\documentclass[12pt]{iopart}
\usepackage{iopams}
\usepackage{graphicx,cite,epsf}
\usepackage{psfrag}
\begin{document}

\title[Shape analysis of random polymer networks]{Shape analysis of random polymer networks}

\author{V. Blavatska$^{1,2}$, K. Haydukivska$^{1,2}$, and Yu. Holovatch$^{1,2,3}$}
\address{$^1$ Institute for Condensed Matter Physics of the National Academy of Sciences of Ukraine,
79011 Lviv, Ukraine}
\address{$^2$ ${\rm L}^4$ Collaboration $\&$ Doctoral College for the Statistical Physics of Complex Systems, Leipzig-Lorraine-Lviv-Coventry, Europe}
\address{$^3$ Centre for Fluid and Complex Systems, Coventry University, Coventry, CV1 5FB, United Kingdom}

\begin{abstract}
We analyze conformational properties of  branched polymer structures, formed on the
base of Erd\"os-R\'enyi random graph model. We consider networks with $N=5$ vertices and
variable parameter $c$, that controls graph connectedness. The universal rotationally invariant size and shape characteristics, such as averaged asphericity
$\langle A_3 \rangle$ and size ratio $g$ of such structures are obtained both numerically by application of Wei's method
 and analytically within the continuous chain model. In particular, our results quantitatively indicate  an increase of asymmetry of polymer network structure
 when its connectedness $c$ decreases.

\end{abstract}

\submitto{Journal of Physics: Condensed Matter}

\section{Introduction}

Size and shape characteristics of individual polymer macromolecules in solvents are
of interest in many aspects. The shape of proteins impacts their folding and motion in the
cellular environment  \cite{Plaxco,Quyang}, the hydrodynamics of polymer solutions
is  affected by the size and shape of individual macromolecules \cite{Torre01}.
In the pioneering work of Kuhn \cite{Kuhn34} it was shown analytically, that flexible polymer chains
in solvents form the crumpled anisotropic coil shapes, resembling rather prolate ellipsoids than spheres. This confirmed  experimental
observations of the viscous properties of polymer solutions, which appeared to differ from those
predicted by the theory of sphere-like molecules \cite{theory}.

\begin{figure}[t!]
	\begin{center}
		\includegraphics[width=130mm]{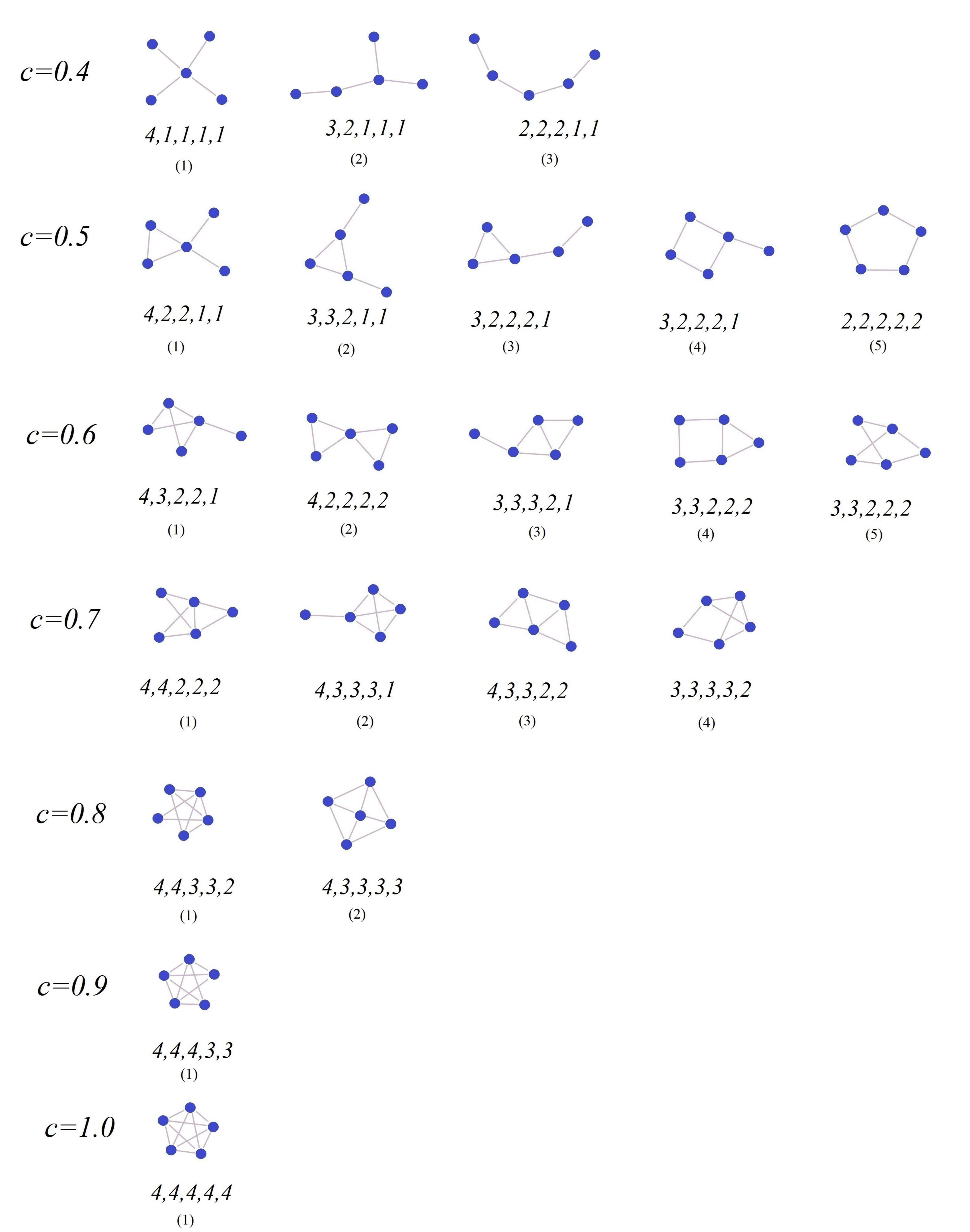}
	\end{center}
	\caption{Schematic presentation of possible configurations of random graphs with $N=5$ vertices and connectedness $c$.
		Only connected configurations are presented. Configurations at fixed $c$ are classified by the set of their vertex degrees, written in descending order below each configuration. For each $c$, one gets $n$ topologically different configurations. 
	}
	\label{network}
\end{figure}

It is convenient to describe and classify the
size and shape properties of polymer configurations in terms of invariants of gyration tensor \cite{Aronovitz86,Rudnick86},
such as the averaged asphericity $ \langle A \rangle$ (here and below,  $\langle \ldots \rangle$ denotes an average
over an ensemble of all possible polymer configurations).
 This quantity equals zero for symmetric spherical configurations,  and takes a value of one
  for completely anisotropic rod-like structures, so that the inequality holds: $0 \leq  \langle A \rangle  \leq 1 $.
An overall size of any polymer structure  can be characterized by the mean-square
radius of gyration $\langle R_g^2 \rangle$.  To compare the size measures of any complex polymer structure $\langle R_g^2 \rangle_{\rm complex}$
and that of the linear polymer chain $\langle R_g^2 \rangle_{\rm chain}$ of the same total molecular weight (same number of monomers), their ratio $g$ is introduced.
This value is universal in the sense, that it is defined only by topological architecture of a complex polymer,
but not by  details of its chemical structure.
Originally, the size ratio   was proposed by Zimm \cite{Zimm49} for  star-like branched polymer structures. In Zimm model, the polymer chain is Gaussian
 and can be considered as a random walk (RW) trajectory.
This approach allows to obtain exact values for polymer conformational characteristics.
Since then, a considerable attention has been paid
to analysis of size and shape characteristics of  macromolecules of linear \cite{Bishop88,Diehl89,Gaspari87,Honeycutt88,Cannon91,Sciutto94,Haber00}, circular \cite{Bishop88,Diehl89,Gaspari87,Cannon91,Jagodzinski92}, and branched \cite{Bishop93,Wei97,Casassa,Ferber13,Ferber15} architectures; in particular the non-trivial
influence of excluded volume on these characteristics has been analyzed.

Hyperbranched polymer structures and polymer networks attract considerable attention  both from
  academic \cite{Duplantier89,Schafer92,Ferber97,Blavatska11} and commercial \cite{Gao04,Jeon18,Gu19} perspective.
Branched polymer networks are the building blocks of materials like synthetic and biological gels, thermoplastics, elastomers.
 The unique properties of these materials such as low viscosity,
high solubility and temperature
stability as well as high functionality
 causes their wide application in fields including drug delivery systems \cite{Li16},
 tissue engineering scaffolds \cite{Lee01}, development of antibacterial/antifouling materials \cite{Zhou10}
 gas storage \cite{McKeown06} etc.


In general, polymer networks can be considered as random structures organized in process of
formation of numerous random chemical bonds (crosslinks) between initial linear polymer chains (strands).
The basis of the polymer network is the junction point, from which linear paths emanate \cite{Dusek69} (a junction is considered as a 
branching point when the number of outgoing strands (functionality) is more than two).
The distribution of junction points in covalent networks is defined by chemical functionalities of
reactant monomers along initial polymer chains.
Some network materials, such as polyurethanes and polyester resins, are often formed on the basis of reactants which carry their reactive
groups (possible junction points) at the ends of chains;
such networks are formed by endlinking polymerization. A network may be formed also by the intermolecular association of polymers, e.g. through hydrogen bonding, producing
 so-called physical networks, such as synthetic and biological gels  \cite{Clark87,Burchard90}.
 The associating polymer (AP) networks are of particular interest in this respect.
An AP is a water-soluble polymer that contains associative
groups (stickers) along its backbone \cite{Winnik97,Chassenieux11}. The stickers on
the chains tend to gather in solvent to form a physically
cross-linked network via the temporal junction of
stickers that has a finite lifetime.

Statistical methods of modeling of polymer network formation
are the question of great interest. These branched
 crosslinked structures  are usually formed by establishing links within the set of polyfunctional
molecules (units) which are called polymer
networks precursors  \cite{Dusek12}. In the language of mathematical graphs, such polymer units can be treated as
vertices, and their chemical functionalities as degrees of these vertices (number of paths eliminating from junction points in polymer networks formed). 
 Starting from the set of polymer units with
  given distribution of  chemical functionalities,  the formation of
 polymer network with known properties results. For example, the polymer networks with a power law distribution of functionalities of junction points (corresponding 
 to the so-called scale-free networks) have been constructed and analyzed recently in Refs. \cite{Dolgushev17,Dolgushev18}.

 Topological characteristics of branched polymer networks  underlying the structure of injectable hydrogels
are important because of their ability to fill cavities with irregular shapes and sizes in  drug delivery systems \cite{Li16}.
Note however, that not much attention has been paid so far to analysis of size and shape characteristics of  hyperbranched complex polymer structures.
 We should mention the results for some deterministic structures of comb-like topologies \cite{Casassa,Ferber13,Ferber15}.
 In the present work, we will turn attention to polymer networks, formed on the base of
 random graph model of Erd\"os and R\'enyi \cite{Renyi}.
 We consider the set of graphs with fixed number of vertices $N=5$ and number of links $L=cN(N-1)/2$ between them.
 The parameter $c$ (changing in range $0 \leq c \leq 1$) is introduced as  graph connectedness,
 taking on a value of $1$ for completely connected graph and $0$ for a set of disconnected vertices
 (Fig. \ref{network}). We will evaluate the size and shape properties of Gaussian polymer structures both numerically by 
 applying the so-called Wei's
  method \cite{Wei} and analytically within the frames of continuous chain model \cite{Edwards}.

  The layout of the rest of the paper is as follows. In the next Section \ref{Def}, we introduce the universal size and shape 
  characteristics of polymer structures.
  In Section \ref{model}, we shortly describe the model of complex polymer architectures we are interested in and present the results, obtained by us by numerically (Subsection \ref{met1})
  and analytically (Subsection \ref{met2}). We end up with Conclusions.

\section{Definitions}\label{Def}

The size and shape characteristics of any polymer structure containing $M$ monomers and embedded in $d$-dimensional space
can be characterized
 \cite{Solc71} in terms of the gyration tensor $\bf{Q}$
with components:
\begin{equation}
Q_{ij}=\frac{1}{M}\sum_{m=1}^M(x_m^i-{x^i_{CM}})(x_m^j-{x^j_{CM}}),\,\,\,\,\,\,i,j=1,\ldots,d.
\label{mom}
\end{equation}
Here, $x_m^i$ is the $i$th coordinate of the position vector $\vec{R}_m$ of $m$th monomer and   ${x^i_{CM}}=\sum_{m=1}^M x_n^i/M$ is $i$th coordinate of the center-of-mass position vector ${\vec{R}_{CM}}$.

 The squared radius of gyration $R_g^2$, which measures the average distribution of monomers with respect to the center of mass,
  is obtained as a trace of the gyration tensor $\bf{Q}$:
\begin{equation}
R_g^2 ={\rm Tr}\, {\bf{Q}} =  \sum_{i=1}^d Q_{ii} =\frac{1}{M}\sum_{m=1}^M (\vec{R}_m-{\vec{R}_{CM}})^2. \label{rg1}	
\end{equation}

 It is established, that the averaged value  $\langle R_g^2 \rangle$, where averaging is performed over an
ensemble of possible configurations of a given polymer structure, scales  with number of monomers $M$ as:
 \begin{equation}
 \langle R_g^2 \rangle \sim M^{2\nu} \label{nu}
 \end{equation}
 with the universal scaling exponent $\nu$ (see e.g. \cite{deGennes,desCloiseaux}). 
 For the case of idealized Gaussian polymer
 one has $\nu=1/2$, whereas taking into account an excluded volume effect leads to $d$-dependence of this exponent.

To compare the size measures of polymer network $\langle R_g^2 \rangle_{\rm network}$
and that of the linear polymer chain $\langle R_g^2 \rangle_{\rm chain}$ of the same total molecular weight (the same number of monomers $M$), the ratio is introduced:
\begin{equation}
g=\frac{\langle R_g^2 \rangle_{\rm network}}{\langle R_g^2 \rangle_{\rm chain}}. \label{gratio}
\end{equation}
Originally, this value was introduced by Zimm \cite{Zimm49} and evaluated for the so-called star polymer architecture - the simplest representative of the class of branched polymer structures, containing one junction point of functionality $f$. For the ideal Gaussian case, this value has been obtained exactly 
\begin{equation}
g_{\rm star}=\frac{3f-2}{f^2}, \label{gratiostar}
\end{equation}
whereas for the Gaussian polymer structure in a form of closed ring one has \cite{Zimm49}:
\begin{equation}
g_{\rm ring}=\frac{1}{2}. \label{gratioring}
\end{equation}
Recently, this ratio has been evaluated \cite{Blavatska15}  
for so-called rosette structure, containing $f_1$ closed rings and $f_2$ linear chains, eliminated from the same branching point:
\begin{equation}
g_{{\rm rosette}}=
\frac{f_1(2f_1-1)+2f_2(3f_2-2)+8f_1f_2}{2(f_1+f_2)^3}. \label{gratiorosette}
\end{equation}
Note,  that the above values (\ref{gratiostar})-(\ref{gratiorosette}) are obtained for the ideal Gaussian polymers without taking into account an excluded volume effect.
In such approach, the exact values for $g$-ratio can be obtained  by e.g. exploiting the so-called continuous chain approach \cite{Edwards}, under the condition that the architecture of polymer structure is not too much complicated.

The spread in the eigenvalues $\sigma_i$ of the gyration tensor (\ref{mom}) describes the distribution of monomers inside the polymer coil and
thus measures an asymmetry of the molecule. For a symmetric (spherical)
configuration all the eigenvalues $\sigma_{i}$ are equal, whereas for the completely stretched rod-like conformation all $\sigma_{i}$ are zero except one.
Let ${\overline{\sigma}}\equiv {\rm Tr}\, {\bf{Q}}/d$
be the  mean arithmetic eigenvalue of  the gyration tensor.
 The shape of polymer structures can be characterized in 	 terms of rotationally invariant quantity called asphericity $A_d$, defined as \cite{Aronovitz86}
 \begin{equation}
 A_d=\frac{1}{d(d-1)}\sum_{i=1}^d\frac{(\sigma_i-{\overline{\sigma}})^2}{{\overline{\sigma}^2}}. \label{asfer1}
\end{equation}

To get experimentally observed average value $\langle A_d \rangle$ one has to perform averaging  
 over an ensemble
of all possible polymer configurations. Note that many analytical studies \cite{Aronovitz86,Rudnick86,Gaspari,Blavatska15}  avoid the averaging of the
ratio in (\ref{asfer1}) due to essential difficulties in calculations and evaluate the quantity:
\begin{equation}
\hat{A_d}=\frac{1}{d(d-1)}\sum_{i=1}^d\frac{\langle (\sigma_i-{\overline{\sigma}})^2 \rangle}{\langle {\overline{\sigma}^2} \rangle}, \label{asfer2}
\end{equation}
where the numerator and denominator are averaged independently
instead of ``truly'' averaged asphericity
 \begin{equation}
\langle A_d \rangle=\left\langle \frac{1}{d(d-1)}\sum_{i=1}^d\frac{(\sigma_i-{\overline{\sigma}})^2}{{\overline{\sigma}^2}} \right\rangle \label{asfer3}.
\end{equation}
However, the latter value can  be  obtained for any complex Gaussian polymer architecture (network)  making use of the Wei method \cite{Wei}, which allows also to estimate the corresponding size ratio (\ref{gratio}).
We describe this method below and apply it to evaluate the size and shape characteristics of the set of polymer structures, constructed on the base of 
Erd\"os-R\'enyi random graph model.

\section{The model and results}\label{model}

The model we propose here describes a set of Gaussian chains randomly linked in junction points, which leads to establishing of a set of multiply branched
 architectures.
Let us start by constructing the graphs with given number of vertices $N$ and number of links $L$ between them, defined by ``connectedness'' $c$.
Parameter $c$ gives the concentration of links as compared with the complete (fully connected) graph (when $L=N(N-1)/2$ and $c=1$). 
The averaged vertex degree is thus given by ${\overline{k}}=\frac{L}{N}=c(N-1)$.
For the case study described in details below, we have considered $N=5$. Such moderate number of vertices allows us to relate our model to real situations at any value of $c$ (and correspondingly at any vertex degree ${\overline{k}}$, which in such case reaches the maximum value of ${\overline{k}}=4$ for the complete graph with $c=1.0$). Indeed, the vertex degree in our description corresponds to the functionality of a junction point in real polymer structure, defined by the chemical functionality of reactive monomers, which normally does not exceed the value of 4 \cite{Burchard90}.  

\begin{figure}[t!]
	\begin{center}
		\includegraphics[width=60mm]{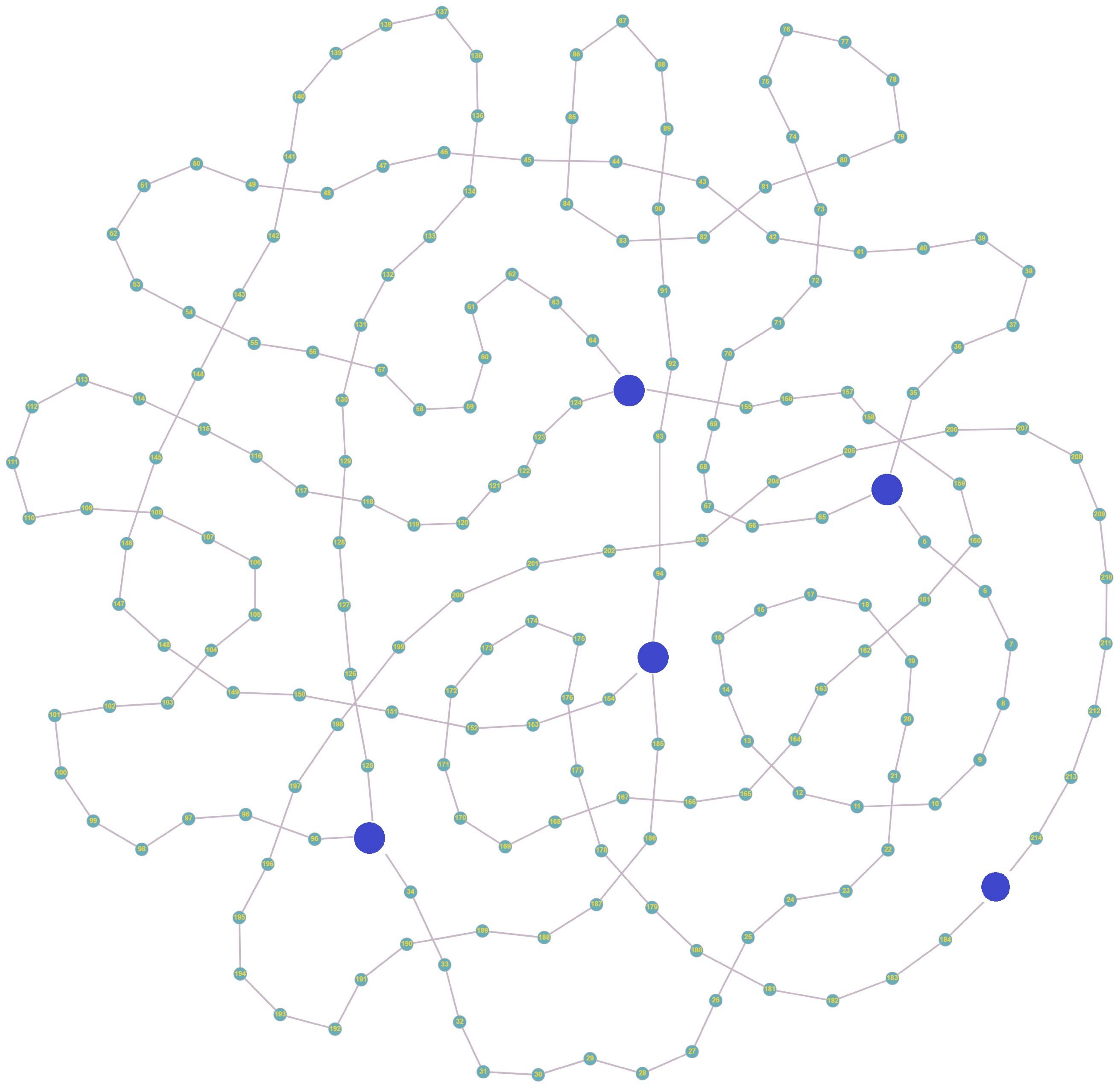}
	\end{center}
	\caption{Schematic presentation of a polymer network, constructed on the basis of the graph  (4) at $c=0.7$
		(as presented in Fig. \ref{network}). Junction points are shown in blue, $l=24$ additional vertices-monomers (shown in green) are inserted along each edge of the  graph. 
		In our analysis, we  analyzed the structures with  number of additional vertices $l$ in every edge up to 40.}
	\label{polymer-network}
\end{figure}

Note, that for the model case $N=5$, concentration $c=0.9$ corresponds to $L=9$, $c=0.8$ to $L=8$ and so on.  We are interested only in all possible connected configurations at given $c$. As a result, we obtain a set of graph configurations, presented in Fig. \ref{network}. At $c<0.4$, no connected graph with $N=5$ vertices can be constructed.
To get  polymer networks, we consider each link in such graphs as a polymer chain with number of monomers (treated as vertices) $l$,
and each vertex with degree $k>1$ as a junction point, so that the resulting graph contains $M=N+L\times l$ vertices.  Schematic presentation of such a polymer network corresponding to  graph (4) at $c=0.7$ with $l=24$ is given in Fig. \ref{polymer-network}.

\subsection{Numerical results: Wei's method}\label{met1}

The size and shape characteristics of branched structures can be computed within the method originally developed by Wei \cite{Wei}.
In general,
Wei's method is applicable to connected network of any topology, if  the
Kirchhoff matrix and its eigenvalues are defined. For the graph of $M$ vertices, Kirchhoff $M\times M$ matrix ${\bf K}$ is defined as follows. Its diagonal elements $K_{ii}$ equal
the degree of vertex $i$, whereas the   non-diagonal elements $K_{ij}$ equal  $-1$ when the vertices $i$ and $j$ are adjacent and $0$ otherwise.
Let $\lambda_2,\ldots,\lambda_M$ be $(M-1)$ non-zero eigenvalues of the $M\times M$
Kirchhoff matrix ($\lambda_1$ is always $0$). Provided that the spectrum $\{ \lambda_i\}$ of matrix ${\bf K}$ is known,  the
asphericity of corresponding Gaussian polymer structure in $d$ dimensions is given by \cite{Wei,Ferber15}:
\begin{equation}
\langle A_d \rangle =\frac{d(d+2)}{2}\int_0^{\infty} {\rm d} y \sum_{j=2}^{M}\frac{y^3}{(\lambda_j+y^2)^2}\left[ \prod_{k=2}^{M} \frac{\lambda_k}{\lambda_k+y^2}\right ]^{d/2},
\label{awei}\end{equation}
whereas the $g$-ratio of the radii of gyration of a network and that of a linear chain with the same number of beads (monomers) is given by:
\begin{equation}
g=\frac{\sum_{j=2}^{M}1/\lambda_j^{{\rm network}}}{\sum_{j=2}^{M}1/\lambda_j^{{\rm chain}}},
\label{gwei}
\end{equation}
where $\lambda_j^{{\rm network}}$ and $\lambda_j^{{\rm chain}}$ are network and chain   
Kirchhoff matrix eigenvalues.

\begin{figure}[t!]
	\begin{center}
		\includegraphics[width=60mm]{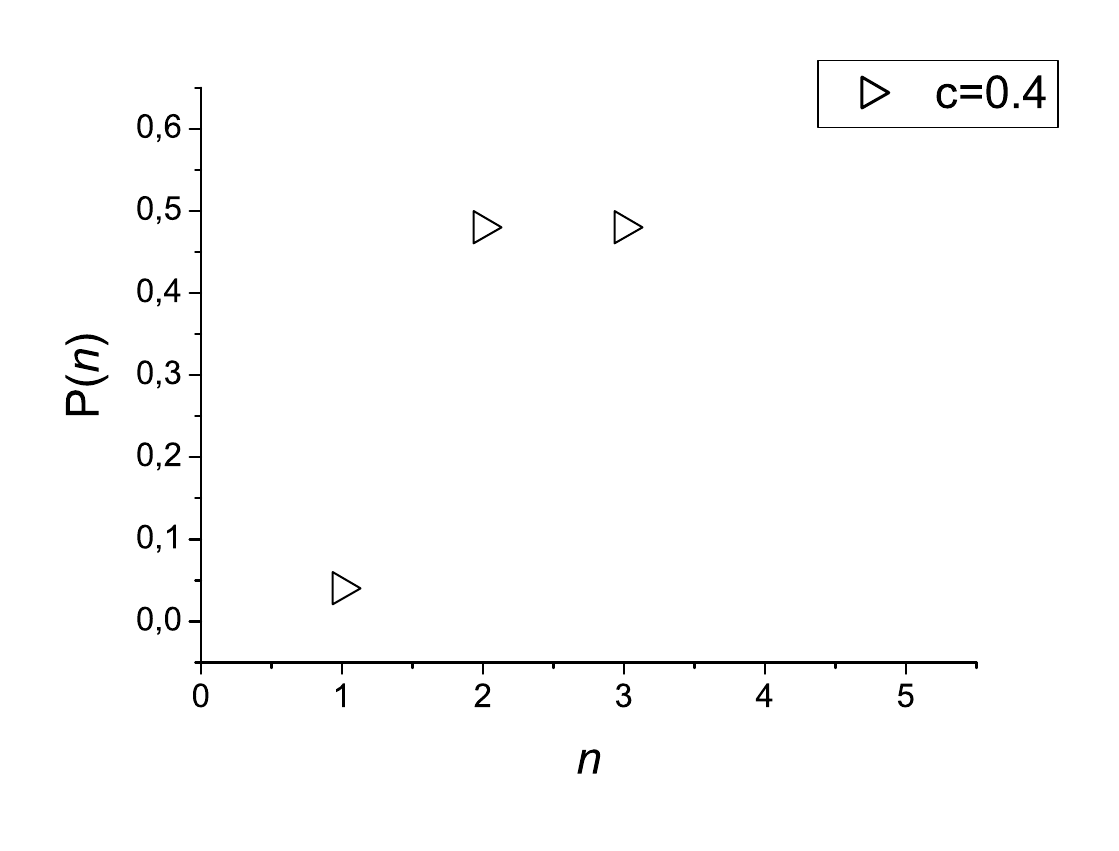}
		\includegraphics[width=60mm]{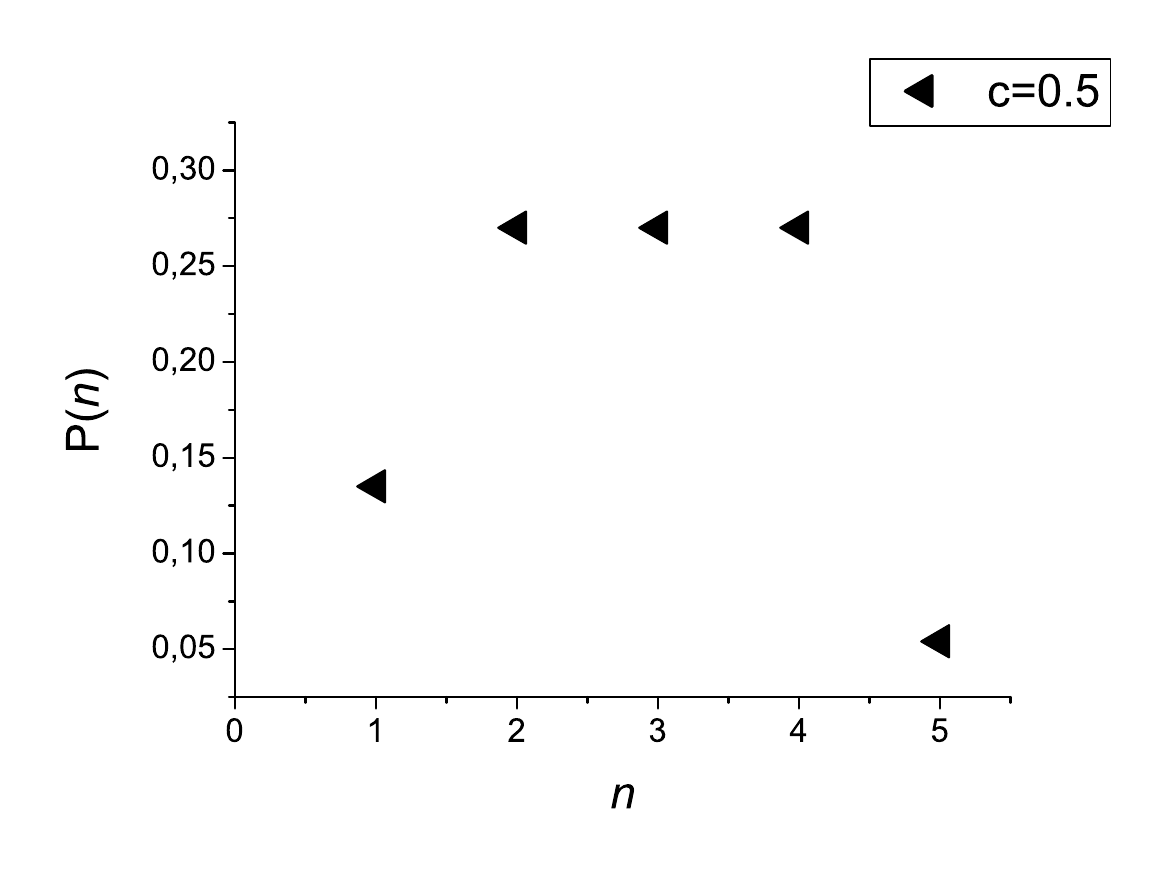}
		\includegraphics[width=60mm]{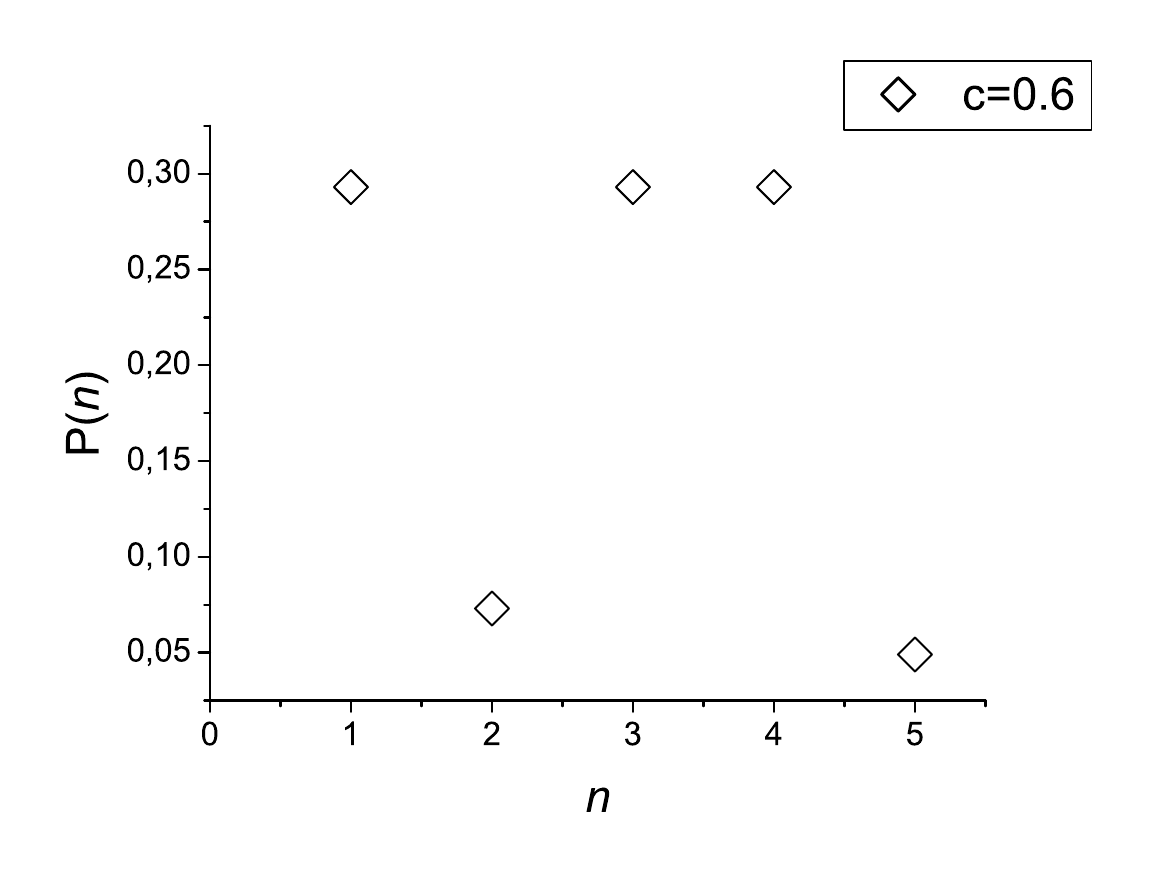}
		\includegraphics[width=60mm]{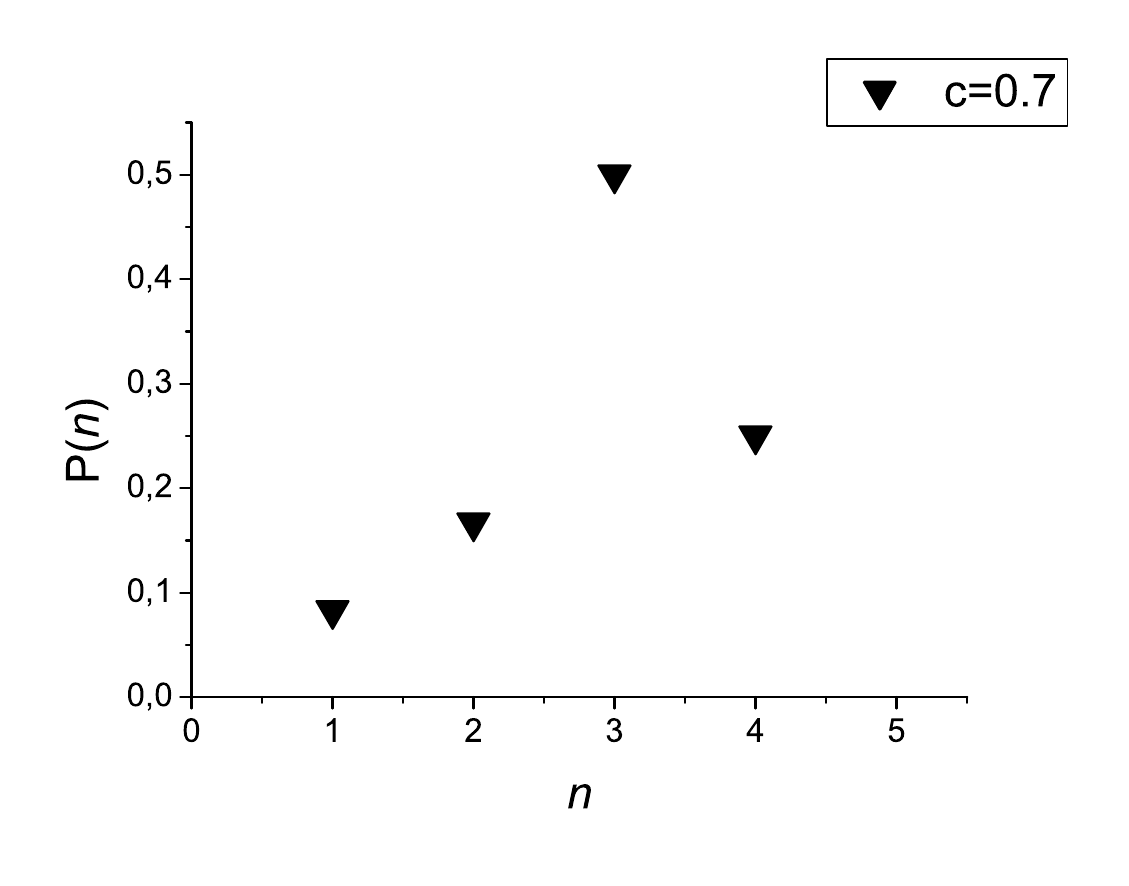}
		\includegraphics[width=60mm]{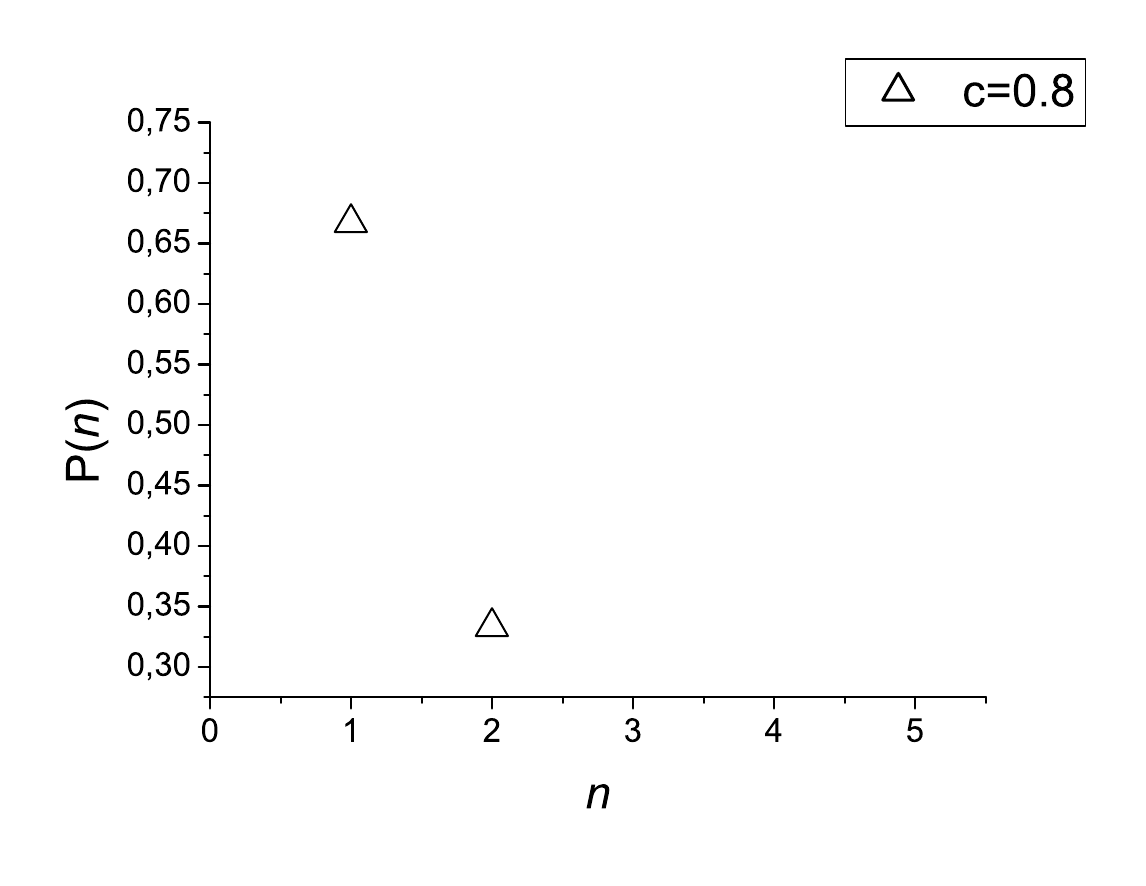}
	\end{center}
	\caption{Probabilities to obtain  graph configurations numbered by $n$ at different $c$. Correspondence between
		number $n$ and graph configuration is shown in Fig. \ref{network}.}
	\label{prob}
\end{figure}

\begin{figure}[t!]
	\begin{center}
		\includegraphics[width=60mm]{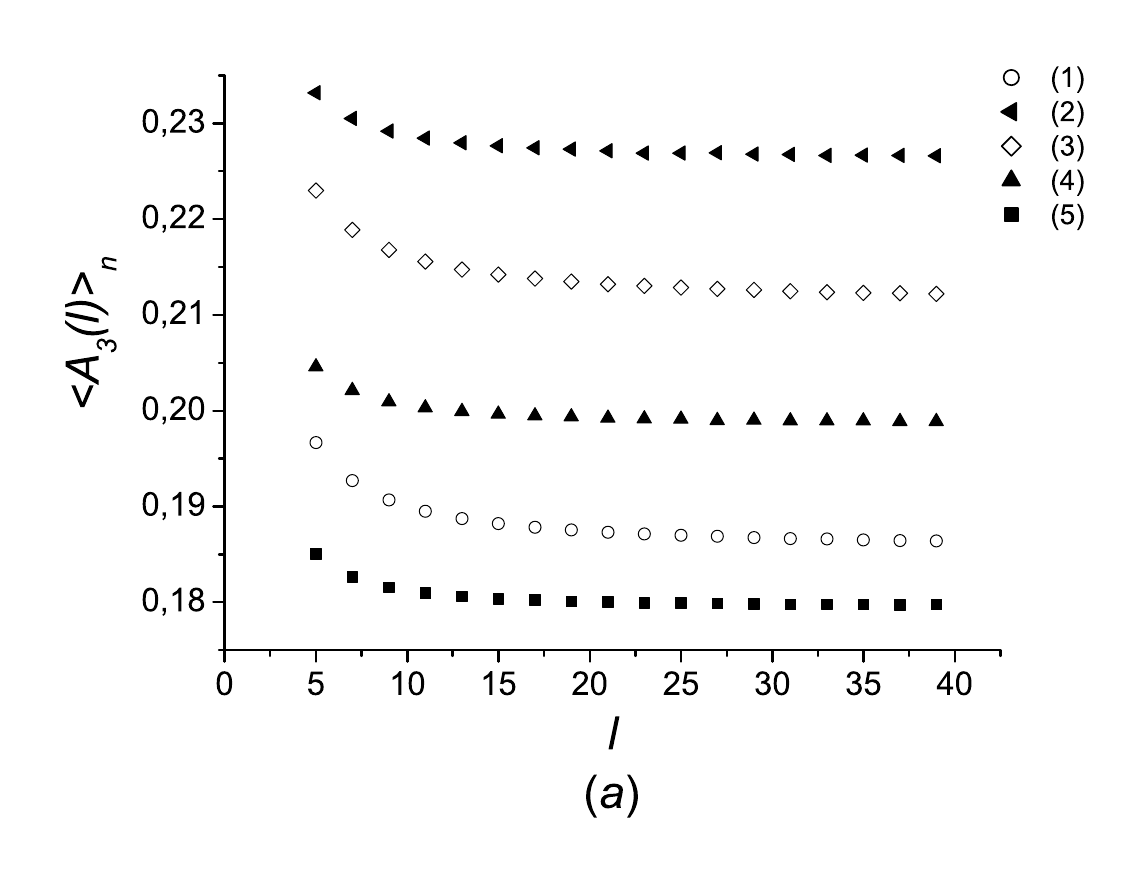}
		\includegraphics[width=60mm]{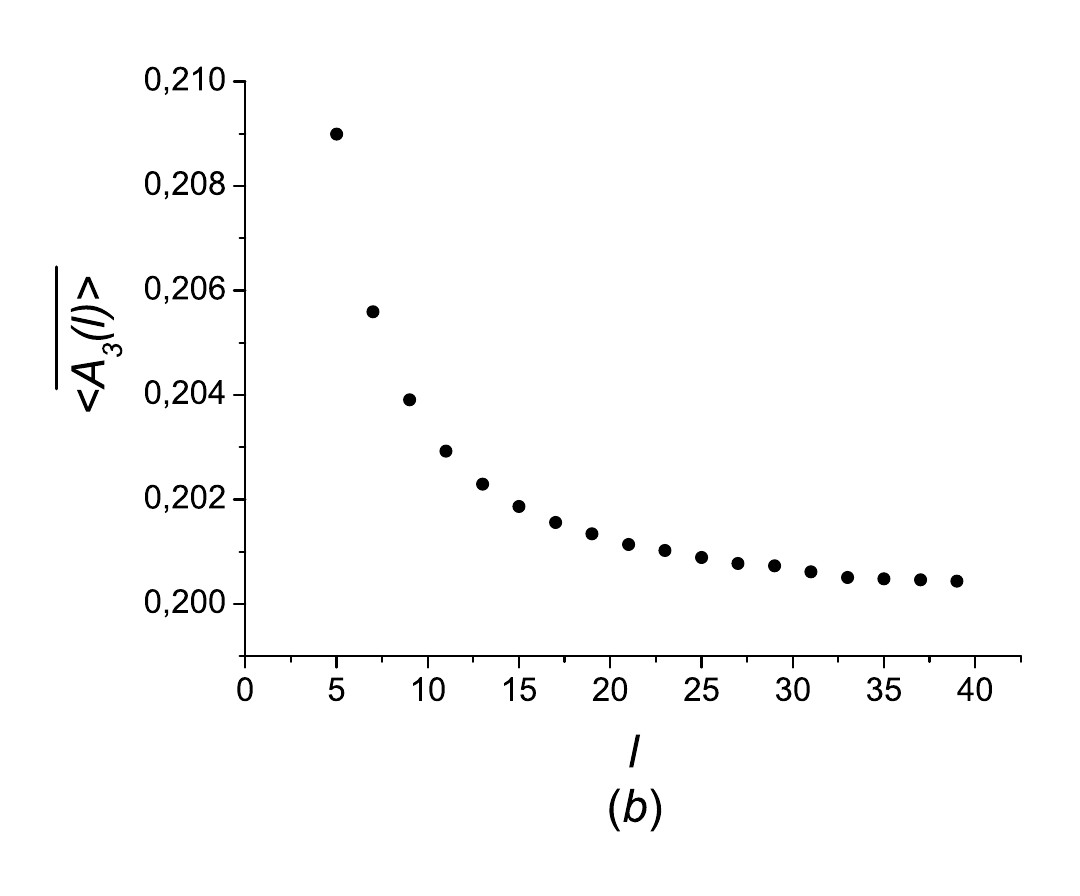}
	\end{center}
	\caption{(a) Individual asphericities $\langle A_3(l) \rangle_n$  of five different polymer network configurations at $c=0.6$, $n=1,\ldots,5$ as shown in Fig. 1, obtained in numerical simulations with application of Wei formula (\ref{awei}), as functions of a single link length $l$. (b) Averaged asphericity ${\overline{\langle A_3 \rangle_n}}$ of an ensemble of possible configurations at $c=0.6$ as a function of a single link length $l$. }
	\label{Asfer-data}
\end{figure}

\begin{figure}[b!]
	\begin{center}
		\includegraphics[width=60mm]{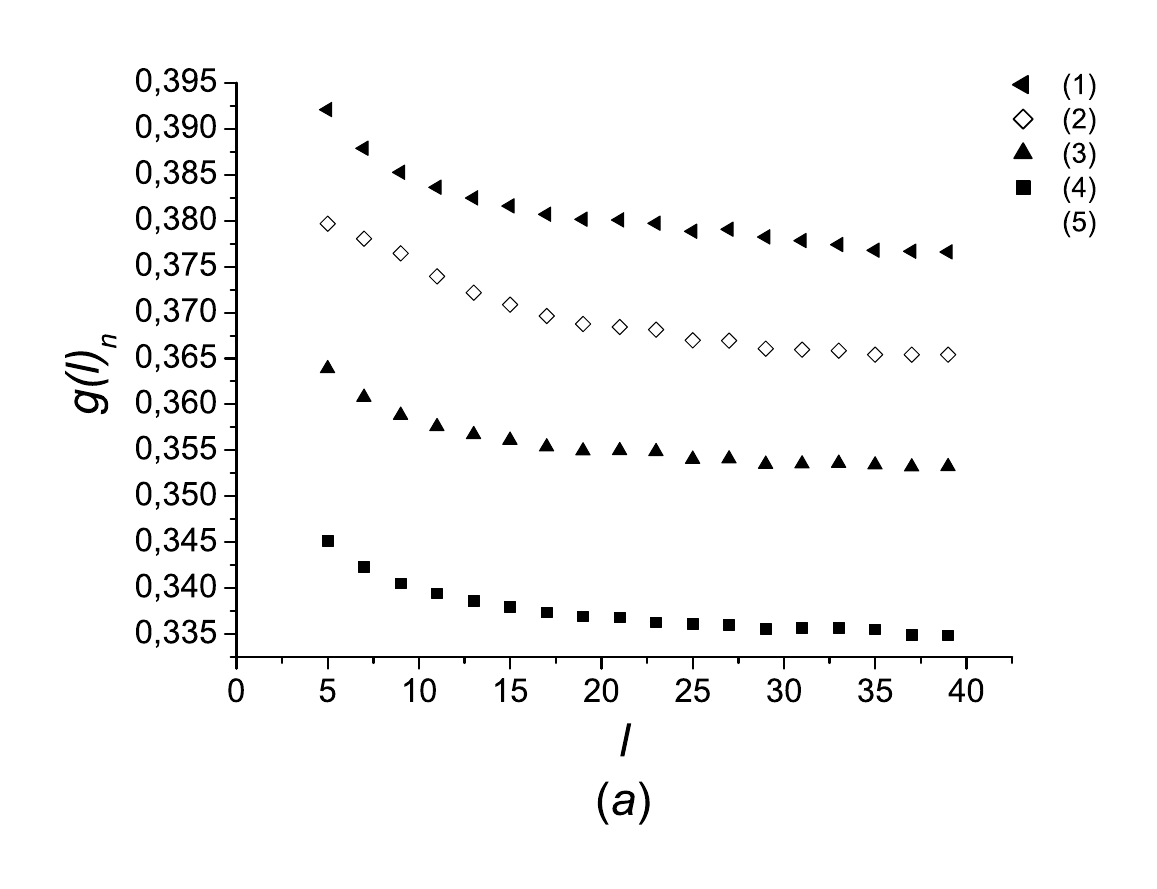}
		\includegraphics[width=60mm]{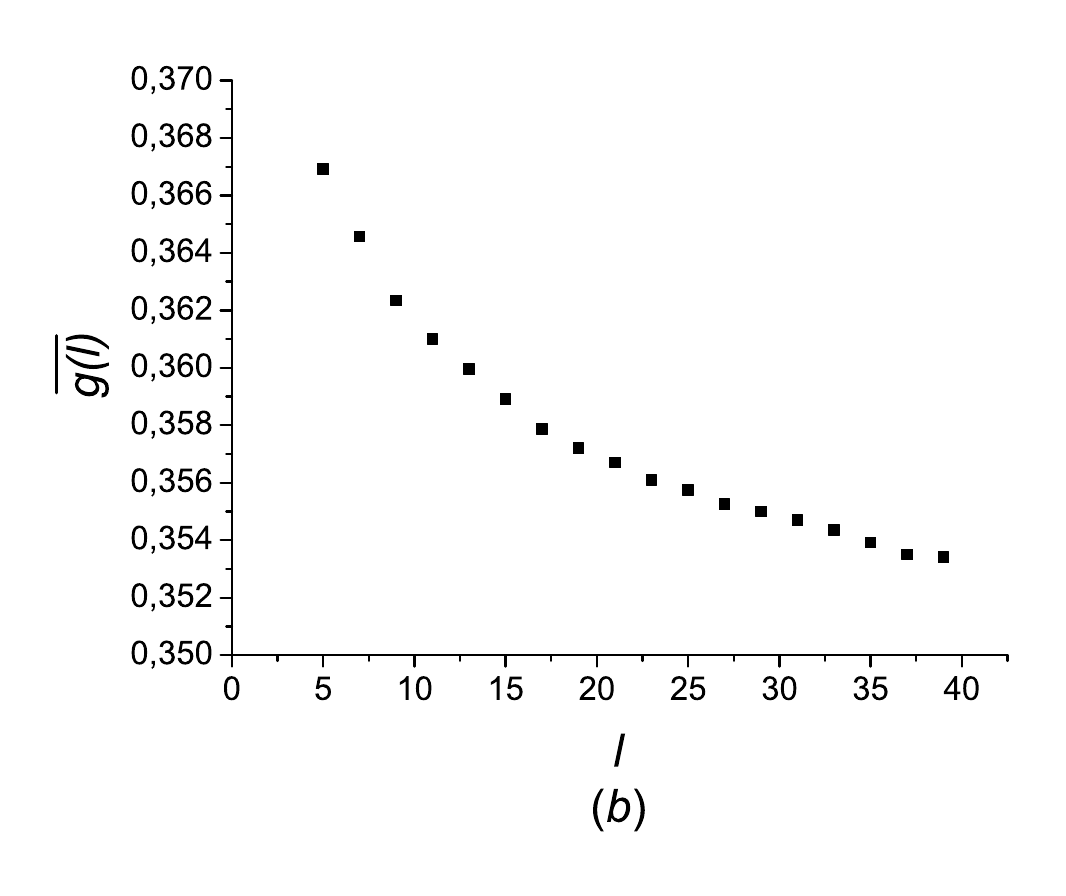}
	\end{center}
	\caption{(a) Individual size ratios $g(l)_n$ of five different polymer network configurations at $c=0.6$, $n=1,\ldots,5$ as shown in Fig. 1, obtained in numerical simulations with application of Wei formula (\ref{awei}), as functions of a single link length $l$. (b) Averaged size ratio $\overline{g}$ of an ensemble of possible configurations at $c=0.6$ as a function of a single link length $l$. }
	\label{g-data}
\end{figure}

In principle, the above parameters are determined within the Wei's approach by extrapolating to infinite length of links $l$.
This is the usual limit of an infinite polymer chain, common in scaling theory of polymer macromolecules \cite{deGennes,desCloiseaux}.
In our analysis, we applied the numerical algorithm of growing chain to analyze the structures with  number of beads $l$ in every link up to 40.
In the rest of analysis presented in this paper we will be interested in the case of polymer networks embedded in three dimensional space, putting $d=3$ in (\ref{awei}), (\ref{gwei}).
 The set of eigenvalues of corresponding Kirchchof matrix was evaluated in each case
and values of $\langle A_3 \rangle$ and $g$ were estimated  according to Eqs. (\ref{awei})-(\ref{gwei}).

Note, that in random process of constructing a graph with the fixed number of vertices and links, each possible configuration
appears with some probability, defined by the number of its realizations. For the fully connected graph with $c=1$, only one realization is possible. For the case $c=0.9$ ($L=9$), we again have only one configuration, though it has 5 realizations, given by the number of possibilities
to choose a pair of vertices with degrees 3. For $c=0.8$ ($L=8$), there are two possible configurations. The first one has 30 possible realizations, the second has 15 realizations. Thus, we can introduce the probabilities for these configurations, $p(1)=30/45=2/3$ and $p(2)=15/45=1/3$ correspondingly.

\begin{table}[t!]
	\begin{tabular}{| c | c | c | c | c |c| c|}
		\hline
		 & $\langle A_3 \rangle_1$ &  $\langle A_3 \rangle_2$ & $\langle A_3 \rangle_3$ & $\langle A_3 \rangle_4$ & $\langle A_3 \rangle_5$ & $\overline{\langle A_3 \rangle}$\\ \hline
		$c=0.4 $ & 0.242(1)  & 0.328(1)  & 0.394(1) & - & -& 0.356(1) \\ \hline
		$c=0.5 $ & 0.224(1)  & 0.246(1)  & 0.320(1)  & 0.241(1) & 0.246(1) & 0.262(1)  \\ \hline
		$c=0.6 $ &  0.185(1) &  0.226(1) & 0.211(1) & 0.198(1) & 0.179(1) & 0.199(1) \\ \hline
		$c=0.7 $ & 0.150(1)   &  0.162(1)  & 0.163(1) & 0.157(1) & - & 0.160(1)  \\ \hline
		$c=0.8 $ & 0.134(1)  &  0.132(1) & - & - & - & 0.133(1)  \\ \hline
		$c=0.9 $ & 0.114(1)  &  - & - & - & - & 0.114(1) \\ \hline
		$c=1.0 $ &  0.099(1) &   &  &  &  & 0.099(1) \\ \hline
		\hline\end{tabular}
	\caption{ Asphericity values $\langle A_3 \rangle_i$ of various polymer network topologies presented in Fig. 1 and their weighted mean value $\overline{\langle A_3 \rangle}$, evaluated on the basis of numerical estimates of Eq. (\ref{awei}).   }\label{table1}
\end{table}

\begin{table}[b!]

	\begin{tabular}{| c | c | c | c | c |c| c|}
		\hline
		 & $g_1$ &  $g_2$ & $g_3$ & $g_4$ & $g_5$ & $\overline{g}$\\ \hline
		$c=0.4 $ & 0.625(1)  & 0.812(1)   & 1 & - & -& 0.895(1) \\ \hline
		$c=0.5 $ & 0.461(1)  & 0.494(1)  & 0.605(1)  & 0.489(1) & 0.500(1)  & 0.518(1)  \\ \hline
		$c=0.6 $ & 0.338(1) & 0.374(1)  & 0.360(1)  & 0.350(1) & 0.332(1) & 0.351(1) \\ \hline
		$c=0.7 $ & 0.259(1)   & 0.267(1)   & 0.268(1) & 0.264(1) & - & 0.267(1)   \\ \hline
		$c=0.8 $ & 0.211(1)  & 0.210(1)  & - & - & - & 0.211(1)  \\ \hline
		$c=0.9 $ & 0.179(1)  &  - & - & - & - & 0.179(1) \\ \hline
		$c=1.0 $ & 0.150(1) &   &  &  &  & 0.150(1) \\ \hline
		\hline\end{tabular}
	\caption{ Size ratio values $g_i$ of various polymer network topologies presented on Fig. 1 and their weighted mean value $\overline{g}$, 
		evaluated on the basis of numerical estimates of Eq. (\ref{gwei}). }\label{table2}
\end{table}

 Following the same way, for $c=0.7$ ($L=7$) we have 4 configurations with numbers of realizations 10, 20, 60 and 30, correspondingly, so that $p(1)=1/12$, $p(2)=1/6$, $p(3)=1/2$, $p(4)=1/4$, and so on.
Probability distributions of graph configurations at various $c$, obtained in our numerical simulations of graph formation, are presented in Fig. \ref{prob}.

Figures \ref{Asfer-data} and \ref{g-data} present examples of  simulation data for  polymer networks, 
constructed on the base of possible configurations of a graph with $c=0.6$ (see the third panel of Fig. 1) 
In Figs. \ref{Asfer-data}a and \ref{g-data}a we show  
individual values 
$\langle A_3(l) \rangle_n$ and $g(l)_n$
for five different polymer networks topologies ($n=1,\ldots,5$).
We evaluate each quantity at fixed link length $l$, calculating spectrum of eigenvalues of Kirchhoff matrices and
 using Wei formulas (\ref{awei}), (\ref{gwei}).
In turn, we plot results as a function of $l$ 
to observe the trend to asymptotic limit. Indeed,
for the finite link length $l$, the values of shape parameters differ from those for infinitely long polymer chains. This finite-size
deviation is usually fitted by:
\begin{eqnarray}
\langle A_3(l) \rangle = \langle A_3 \rangle + a/l,\nonumber\\
 g(l)  =  g  + b/l,\label{fit}
\end{eqnarray}
where $a$, $b$ are constants.
The shape parameter estimates can be obtained by least-square fitting of (\ref{fit}). The results are presented in Tables \ref{table1}, \ref{table2}.

Remembering, that any graph configuration $n$ on Fig. \ref{network} is realized with some probability $p(n)$ (see Fig. \ref{prob}), we may propose the following interpretation for real polymer systems. Let us consider the mixture of associative polymer chains with functional junction points, able to establish mutual links. Assume, that we are interested  in complex polymer structures, formed in such a mixture, with fixed number of junction points $N=5$ and some fixed number of polymer chains included (corresponding to parameter $L$). This will correspond to
considering with certain probability $p(n)$ possible configurations numbered by $n$ at fixed $c$, presented in Fig. (\ref{network}). Thus, we may be interested in averaged values of asphericity and size ratio at fixed $c$, given by:
 \begin{eqnarray}
&& {\overline{\langle A_3 \rangle}}=\sum_n p_n \langle A_3\rangle_n,\,\,\,
  {\overline{g}}=\sum_n p_n  g_n, \label{aver}
 \end{eqnarray}
where $\langle A_3\rangle_n$, $g_n$ are corresponding values of individual configuration $n$. Resulting values at each $c$ are presented in the last columns of Tables \ref{table1}, \ref{table2}.
We note, that the averaged asphericity and thus the degree of asymmetry of typical polymer configuration increases with decreasing the connectedness $c $ of corresponding graphs.
Also, the size ratio $g$ increases and tends towards the limiting value of $1$ with decreasing parameter $c$. At large $c$ values, the configurations of polymer structures are more compact in space, whereas at smaller $c$ they become more elongated and resemble linear polymer chain of the same total molecular weight.

Note, that some of graph configurations in Fig. \ref{network} restore the polymer architectures, already studied before, for which the exact values of the size ratio $g$ are known.
 In particular, the configuration (1) at $c=0.4$ restores the star polymer, and its size ratio is given by Eq. (\ref{gratiostar}) with $f=4$: $ g=5/8$,  whereas (3) at $c=0.4$ is a simple linear topology with $g=1$. Configuration (5) at $c=0.5$ restores the ring polymer architecture with the size ratio given by (\ref{gratioring}): $g=1/2$. Configuration (2) at $c=0.6$ is a representative of rosette polymer architecture, whose size ratio is given by (\ref{gratiorosette}) with $f_1=2$, $f_2=0$: $g=3/8$. Note that the numbers quoted in Table \ref{table2} reproduce their exact counterparts with nice accuracy. Similar accuracy is expected for the rest of 
the values given in the Table 
where the exact results have not been obtained so far.

Whereas Wei's method can be easily applied to evaluate the shape characteristics of Gaussian polymer structure of
any topology, an analytical approach based on continuous chain model
 allows to derive the {\it exact} values of these characteristics. However, the latter method encounters limitations caused by cumbersome calculations with growing complexity of polymer architecture. In the next Subsection, we apply the continuous chain model to evaluate the size ratio $g$ of some structures depicted on Fig. \ref{network} in order to compare them with our numerical results.

\subsection{Analytical results: continuous chain model}\label{met2}

\begin{figure}[b!]
\begin{center}
\includegraphics[width=150mm]{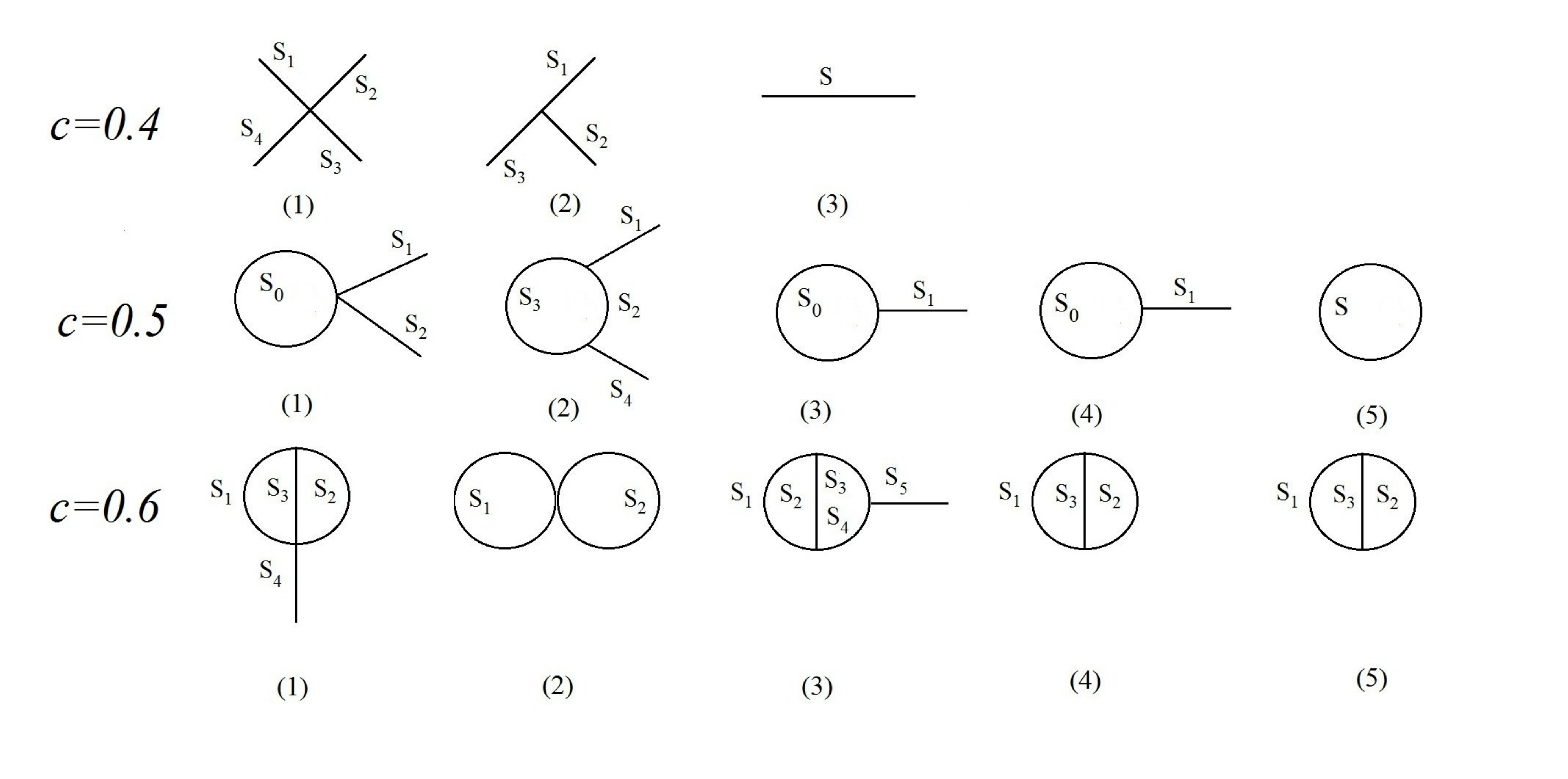}
\caption{ \label{diagramy6} Schematic presentation of polymer structures, corresponding to graphs shown on Fig. \ref{network} within the frames of the continuous chain model. Each line denotes a path of length $S_i$.}
\end{center}
\end{figure}

Another approach that can be used to describe and analyze  the polymer structures presented in Fig. \ref{network} is based on the continuous chain model proposed by Edwards \cite{Edwards}. Within this model a polymer chain is presented as a trajectory of a certain length $S$, that can also be associated with the number of monomers  \cite{desCloiseaux}. This parameter corresponds to parameter $l$ in previous Section. The trajectory can be parametrized by a radius vector $\vec{r}(s)$,
 with $0\leq s \leq S$ \cite{Edwards}. The Hamiltonian of a single trajectory in the Gaussian case is given by:
\begin{eqnarray}
 H=\frac{1}{2}\int_0^{S}\left( \frac{{\rm d}\vec{r}(s)}{{\rm d}s}\right)^2, \label{H}
\end{eqnarray}
which takes into account the connectedness of the trajectory, the latter is allowed to cross itself.
The  partition function (number of configurations) of a single linear polymer chain reads:
\begin{eqnarray}
&&Z(S) = \int\,D\vec{r}(s) {\rm e}^{-\frac{1}{2}\int\limits_0^{S}
	\left( \frac{ {\rm d}\vec{r}(s)}{{\rm d}s}  \right)^2
},\label{Z}
\end{eqnarray}
here $\int\,D\vec{r}(s)$ denotes functional path integration over all possible trajectories that start at the origin and are of  length $S$.

Within the framework of this model it is possible to calculate some of the universal conformational properties of polymer structures. For example, the 
mean-squared gyration radius as defined by (\ref{rg1}), within the frames of the continuous model is given by:
\begin{eqnarray}
&&\langle R_{g}^2 \rangle = \frac{1}{2S^2} \left\langle  \int_0^{S} \!\! d z_1\int_0^{ S} \!\! d z_2
(\vec{r}( z_2)-\vec{r}( z_1 ))^2 \right\rangle \label{rg}.
\end{eqnarray}
Here,  $z_1$ and $z_2$ the so called restriction points \cite{desCloiseaux} and $ \langle ( \ldots) \rangle $ denotes an averaging over all possible trajectories with the Hamiltonian (\ref{H}):
\begin{eqnarray}
\langle  (\ldots) \rangle = \frac{\int\,D\vec{r}(s)\, {\rm e}^{-H} (\ldots)}{Z(S)}. \label{sered}
\end{eqnarray}

Peculiarities of architecture of any complex polymer structure, containing branching points and closed loops, can be taken into account in the partition function by introducing the corresponding set of $\delta$-functions \cite{Duplantier89}. Also, one has to take into account that the linear strands within such complex structures
can be treated as trajectories of various $S_i$. 
The simplest polymer architectures are presented by structures at $c=0.4$ (see Fig. \ref{network}), for which a certain number $f$ of strands starts from the common branching point (the so-called star-like polymer) \cite{Zimm49}. The corresponding partition function can be presented as:
\begin{eqnarray}
&&Z_1(\{S_i\}) = \int\,D\vec{r}(s)\,\prod_{i=1}^{f}\delta(\vec{r_i}(0)) \exp(-\sum_{i=1}^f H_i), \label{Z1}
\end{eqnarray}
 where $S_i$ is a length of $i$th strand,  $H_i$ being an effective Hamiltonian of the $i$th strand:
 \begin{equation}
  H_i=\frac{1}{2}\int_0^{S_i}\left( \frac{{\rm d}\vec{ r}_i(s)}{{\rm d}  s}\right)^2.
\end{equation}
The product of $\delta$-functions in (\ref{Z1}) ensures the star-like structure of a set of $f$ polymer chains: all of them start at the same core point $\vec{r_i}(0)=0$. Keeping the parameters $S_i$ different allows us to  
describe chains of different lengths, as further explained below.
 
Let us start with the case $c=0.4$ in Fig. 1. We notice, that within the continuous chain model,  the corresponding structures can be described as follows: structure (1) contains four strands of equal length; structure (2) contains two strands of equal length and one strand twice larger; structure (3) corresponds to a single linear strand.   
  Let us introduce $S=\sum_{i=1}^4 S_i$ as the total length. Thus, structure (1) corresponds to the star-like polymer with $f=4$ and all strands of equal lengths $S_i=S/4$, structure (2) corresponds to $f=3$ strands,  one of them is two times longer than the others ($S_1=S_2=S/4$, $S_3=2S_1=S/2$), and structure  (3) is a single chain ($f=1$) of length $S$.
 In the upper panel of Fig. \ref{diagramy6}  we give schematic representation of these structures. 

The structures (1), (3)-(5) in Fig. \ref{network} with $c=0.5$  contain a loop of certain length $S_0$ and $f$ linear strands expanding from the same junction point, so that $S_0+\sum_{i=1}^f S_i=S$. In general, the partition function of such structures is given by:
\begin{eqnarray}
&&Z_2(\{S_i\}) = \int\,D\vec{r}(s)\,\prod_{i=1}^{f}\delta(\vec{r_i}(0))\,\delta(\vec{r_0}(S_0)-\vec{r_0}(0))\exp(-\sum_{i=0}^{f}H_i). \label{Z2}
\end{eqnarray}
Here, the second $\delta$-function ensures the closed loop structure of the $0$th strand: it starts and ends at the same point $\vec{r_0}(S_0)=\vec{r_0}(0)$.
Thus, structure (1) at $c=0.5$ corresponds to $f=2$ and $S_1=S_2=S/5$, $S_0=3S_1=3S/5$; structures (3) and (4) contain single linear strand ($f=1$) with lengths correspondingly $S_1=2S/5, S_0=3S/5$ and $S_1=S/5,S_0=4S/5$; and structure (5) is a single loop (ring polymer) of length $S_0=S$, $f=0$ (see the middle panel of Fig. \ref{diagramy6}).

The remaining structure (2) at $c=0.5$ is considered separately since it contains two branching points. Corresponding partition function reads:
\begin{eqnarray}
&&Z_3(\{S_i\}) = \int\,D\vec{r}(s)\,\prod_{i=1}^{3}\delta(\vec{r_i}(0))\delta(\vec{r_3}(S_3)-\vec{r_2}(S_2))\times \nonumber\\
&&\times\delta(\vec{r_3}(S_3)-\vec{r_4}(0)) \exp(-\sum_{i=0}^{4}H_i),
\end{eqnarray}
where $S_1=S_2=S_4=S/5,S_3=2S/5$. The function $\delta(\vec{r_3}(S_3)-\vec{r_2}(S_2))$ describes the fact that trajectories number 
two and three end at the same point,  and $\delta(\vec{r_3}(S_3)-\vec{r_4}(0))$ ensures that the fourth trajectory starts at that point too.

Finally, let us consider the case $c=0.6$ in Fig. 1. Corresponding structures in continuous chain model representation are given in the lower panel of Fig. \ref{diagramy6}.  The structure (1) is described by the partition function:
\begin{eqnarray}
&&Z_4(\{S_i\}) = \int\,D\vec{r}(s)\,\prod_{i=1}^{3}\delta(\vec{r_i}(0))\,\delta(\vec{r_3}(S_3)-\vec{r_1}(S_1))\,\delta(\vec{r_3}(S_3)-\vec{r_2}(S_2))\times\nonumber\\
&&\times\delta(\vec{r_3}(S_3)-\vec{r_4}(0))\exp(-\sum_{i=1}^{4}H_i), 
\end{eqnarray}
 with $S_1=S_2=S/3$, $S_3=S_4=S/6$.
  The structure (2) contains two connected loops and is described by:
 \begin{eqnarray}
 &&Z_5(\{S_i\}) = \int\,D\vec{r}(s)\,\prod_{i=1}^{2}\delta(\vec{r_i}(0))
 \,\delta(\vec{r_1}(S_1)-\vec{r_2}(S_2))\exp(-\sum_{i=1}^{2}H_i), 
 \end{eqnarray}
 with  $S_1=S_2=S/2$.
The partition function for structure (3) reads:
\begin{eqnarray}
&&Z_6(\{S_i\}) = \int\,D\vec{r}(s)\,\prod_{i=1}^{3}\delta(\vec{r_i}(0))\delta(\vec{r_1}(S_1)-\vec{r_2}(S_2))\delta(\vec{r_3}(S_3)-\vec{r_4}(S_4))\times\nonumber\\
&&\times\delta(\vec{r_2}(S_2)-\vec{r_4}(0))\delta(\vec{r_3}(S_3)-\vec{r_5}(0))\exp(-\sum_{i=1}^{5}H_i)\nonumber\\
\end{eqnarray}
with $S_1=S/3,S_2=S_3=S_4=S_5=S/6$.
Structures (4) and (5) look identically and are described by the partition function
\begin{eqnarray}
&&Z_7(\{S_i\}) = \int\,D\vec{r}(s)\,\prod_{i=1}^{3}\delta(\vec{r_i}(0))\,\delta(\vec{r_3}(S_3)-\vec{r_1}(S_1))\,\delta(\vec{r_3}(S_3)-\vec{r_2}(S_2))\times\nonumber\\
&&\times \exp(-\sum_{i=1}^{3}H_i). \label{Z7}
\end{eqnarray}
For structure (4) one should take  $S_1=2S/3, S_2=S/3,S_3=S/6$, whereas for structure (5): $S_1=S_2=S/3,S_3=S/3$.

\begin{figure}[t!]
\begin{center}
\includegraphics[width=80mm]{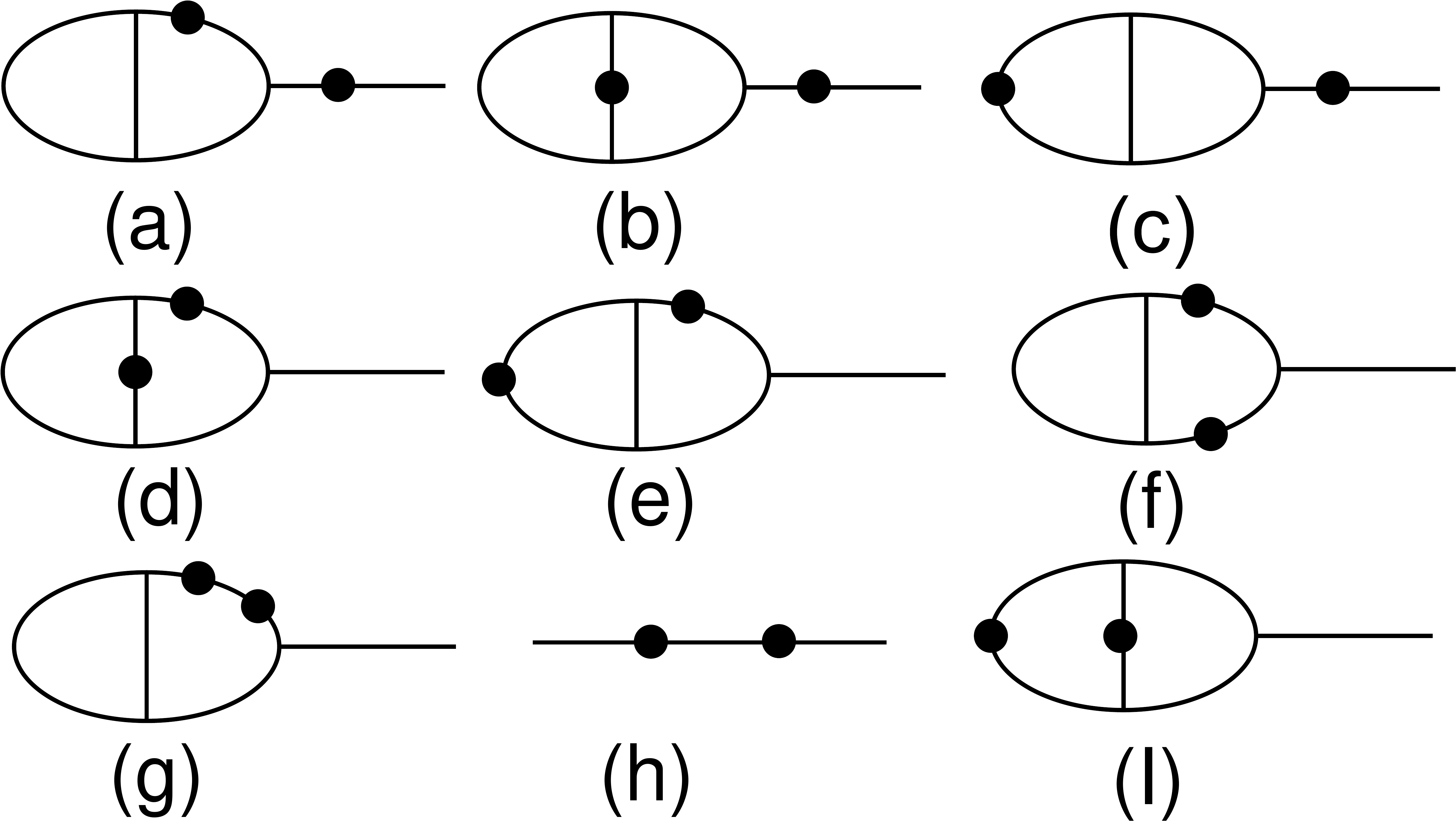}
\caption{ \label{rgdiagram} Diagrammatic presentation of contributions into $\langle \xi(\vec{k}) \rangle$  of the polymer structure (3) at $c=0.6$ (as shown in Fig. \ref{diagramy6}). The solid line on a diagram is a schematic presentation of a polymer path of length $S_i$, and
bullets denote the so-called restriction points $ z_1$ and $ z_2 $.}
\end{center}
\end{figure}

\begin{table}[b!]
	\begin{tabular}{| c | c | c | c | c |c| c|}
		\hline
		 & $g_1$ &  $g_2$ & $g_3$ & $g_4$ & $g_5$ & $\overline{g}$\\ \hline
		$c=0.4 $ & 5/8  & 13/16   & 1 & - & -& 179/200 \\ \hline
		$c=0.5 $ & 23/50  & 123/250  & 151/250  & 61/125 & 1/2  & 957/1850   \\ \hline
		$c=0.6 $ & 97/288 & 3/8  & 13/36  & 139/396 & 1/3 & 949/2706 \\ \hline
		\hline\end{tabular}
	\caption{ Our results for the size ratios of various polymer network topologies presented in Fig. 1, obtained within the continuous chain model }\label{table3}
\end{table}

As we have already mentioned above, although the continuous chain model enables description of polymers of any architecture, in practice the calculations become cumbersome with growing complexity of the polymer architecture. Therefore, below we display results only for the polymer architectures described by the partition functions  $Z_1(\{S_i\})$-$Z_7(\{S_i\})$, Eqs. (\ref{Z1}), (\ref{Z2})-(\ref{Z7}) and shown in Fig. \ref{diagramy6}.

To evaluate the gyration radii of polymer structures described above, we generalize the definition (\ref{rg}) as follows:
\begin{eqnarray}
&&\langle R_{g,n}^2\rangle = \frac{1}{2S^2} \left\langle \sum_{i,j=1}^{f_n} \int_0^{S_i} \!\! d z_1\int_0^{ S_j} \!\! d z_2
(\vec{r}_i( z_2)-\vec{r}_j( z_1 ))^2 \right\rangle, \,\,\,n=1,\ldots,7. \label{rgn}
\end{eqnarray}
Here $n$ numerates the structures classified by the partition functions as given by Eqs. (\ref{Z1}), (\ref{Z2})-(\ref{Z7}),  $f_n$ is the number of strands in the corresponding structure and $S=\sum_{i=1}^{f_n} S_i$ is the total length.
 The radius of gyration  (\ref{rgn})  can be calculated 
using the identity 
\begin{equation} 
(\vec{r}_i( z_2)-\vec{r}_j( z_1 ))^2=-2{ d}\frac{{\partial}}{{\partial} |k|^2} \xi(\vec{k}) \Big|_{k=0},\,\,\,\xi(\vec{k})\equiv{\rm e}^{i\vec{k}\vec{r}_i( z_2)-\vec{r}_j( z_1 )},\label{def} 
\end{equation} 
with $d$ being the space dimension,
and evaluating $\langle \xi(\vec{k}) \rangle$ in the path integration approach.   
In calculation of the contributions into $\langle \xi(\vec{k}) \rangle$ it is convenient  to use
the diagrammatic presentation as given in Fig. \ref{rgdiagram}. 
According to 
the general rules of diagram calculations \cite{desCloiseaux}, each segment between any two restriction points $z_a$ and $z_b$ 
bears a wave vector $\vec{q}_{ab}$.  It is given by a sum of incoming and outcoming  wave vectors, which are injected 
at corresponding restriction points. 
At these points, the flow of wave vectors is 
conserved. 
A factor $\exp\left(-{\vec{q}_{ab}}^{\,\,2}(z_b-z_a)/2\right)$ is associated with each segment. An integration is performed 
over all independent segment areas and over wave vectors injected at the end points. An example of diagram evaluation is given in the Appendix.

As a result, we obtained expressions for the gyration radii $\langle R^2_{g,n}\rangle$, corresponding to partition functions $Z_n(\{S_i\}$ above:
\begin{eqnarray}
&&\langle R^2_{g,1}\rangle= \frac{d}{(\sum_{i=1}^{f} S_i)^2}\left( \sum_{i=1}^f\frac{S_i^3}{6}+\sum_{j\neq i=1}^f\frac{S_i^2S_j}{2}\right),\\
&&\langle R^2_{g,2}\rangle= \frac{d}{(S_0+\sum_{i=1}^{f}S_i)^2}\left(\frac{S_0^3}{12}+\nonumber\right.\\
&&\left.+\sum_{i=1}^f\left(\frac{S_i^3+S_iS_0^2}{6}+\frac{S_i^2S_0}{2}+\sum_{j\neq i=1}^f\frac{S_i^2S_j}{2}\right)\right),\\
&&\langle R^2_{g,3}\rangle= \frac{d}{12(\sum_{i=1}^4 S_i)^2(S_2+S_3)}\left(2(S_1+S_4)^3(S_3+S_2)+\right.\nonumber\\
&&\left.+6(S_3+S_2)^2(S_1^2+S_4^2)\right.\nonumber\\
&&\left.+12S_1S_2S_3S_4+(S_3+S_2)^4+2(S_3+S_2)^3(S_1+S_4)\right),\label{R3}\\
&&\langle R^2_{g,5}\rangle= \frac{d(S_1+S_2)}{8},\\
&&\langle R^2_{g,6}\rangle= \frac{d}{(\sum_{i=1}^5 S_i)^3}\left(\frac{S_1^3}{12}+\frac{S_2^3}{12}\frac{(S_3+S_4)^3}{12}+\frac{S_5^3}{6}+\right.\nonumber\\
&&\left.+\frac{1}{6}(S_3+S_4)^2(S_1+S_2)+\frac{1}{6}(S_3+S_4)(S_1^2+S_2^2)\right.\nonumber\\
&&+\frac{1}{6}(S_1^2S_2+S_1S_2^2)+\frac{S_5^2}{2}(S_1+S_2+S_3+S_4)+\nonumber\\
&&+\frac{S_1S_2(S_3+S_4)(S_1+S_2+S_3+S_4)^2}{12(S_1S_2+(S_1+S_2)(S_3+S_4))}\nonumber\\
&&+\frac{S_5(S_1S_2+S_3S_4)(S_1+S_2)(S_3+S_4)}{2(S_1S_2+(S_1+S_2)(S_3+S_4))}+\nonumber\\
&&+\frac{S_5(S_1S_2(S_3^2+S_4^2)+2S_3S_4(S_1+S_2)^2)}{2(S_1S_2+(S_1+S_2)(S_3+S_4))}\nonumber\\
&&\left.+\frac{S_5((S_1^3+S_2^3)(S_3+S_4))+S_1^3S_2+S_1S_2^3}{6(S_1S_2+(S_1+S_2)(S_3+S_4))}\right),\\
&&\langle R^2_{g,7}\rangle= \frac{d}{(\sum_{i=1}^3 S_i)^2}\left(\sum_{i=2}^3\left(\frac{S_i^3}{12}+
\frac{S_iS_1^2}{2}+\sum_{j\neq i=1}^3\frac{S_i^2S_j}{6}\right)+\right.\nonumber\\
&&\left.+\frac{S_2S_3S_1(S_2+S_3+S_1)^2}{12(S_2S_3+S_2S_1+S_3S_1)}\right).
\end{eqnarray}

To evaluate the size ratios, defined for each structure by (\ref{gratio}), we divide corresponding $\langle R^2_{g,n}\rangle$ from expressions above by the value 
of the gyration radius 
of a linear polymer chain with the same total length $S$ given by
$\langle R^2_{g}\rangle_{chain}=\frac{Sd}{6}
$. 
As an example, let us consider the structure (2) at $c=0.5$, with the gyration radius given by Eq. (\ref{R3}).
Substituting $S_1=S_2=S_4=S/5$, $S_3=2S/5$ into $\langle R^2_{g,3}\rangle$, we obtain the ratio:
\begin{equation}
\frac{\langle R^2_{g,3}\rangle}{\langle R^2_{g}\rangle_{chain}}=\frac{123}{250}.
\end{equation}

Note that for the structure (1) at $c=0.4$ we reproduce the known value of the size ratio of a star polymer as given by Eq. (\ref{gratiostar}) at $f=4$:
\begin{equation}
\frac{\langle R^2_{g,1}(f=4,S_i=S/4)\rangle}{\langle R^2_{g}\rangle_{chain}}=\frac{5}{8},
\end{equation}
for structure (5) at $c=0.5$ the size ratio (\ref{gratioring}) is obtained:
\begin{equation}
\frac{\langle R^2_{g,2}(f=0,S_0=S,S_1=S_2=0)\rangle}{\langle R^2_{g}\rangle_{chain}}=\frac{1}{2},
\end{equation}
whereas structure (2) at $c=0.6$ restores rosette structure with the size ratio given by Eq. (\ref{gratiorosette})  at $f_1=2$, $f_2=0$:
\begin{equation}
\frac{\langle R^2_{g,5}\rangle}{\langle R^2_{g}\rangle_{chain}}=\frac{3}{8}.
\end{equation}

The exact results for the size ratios for the rest of the structures were unknown so far, we summarize all of them in table \ref{table3}. 
Also we note that for all  structures under consideration the analytic results are in perfect agreement with the numerical data, obtained within the frames of Wei's method as presented in the previous subsection.

\section{Conclusions}\label{conc}

In this paper, we proposed the model of a random  polymer network, formed on the base on 
 Erd\"os-R\'enyi random graph \cite{Renyi}. 
 In the language of mathematical graphs, the chemical bonds between monomers (junction points of networks) 
  can be treated as
vertices, and their chemical functionalities as degrees of these vertices.
 Being sufficiently simple, such a model captures main features of polymer networks: 
 their connectivity and relatively small values of node degrees that correspond to 
chemical valencies of separate monomers.
 
 
 We consider  graphs with fixed number of vertices $N=5$ and variable parameter $c$ (``connectedness''), defining the total number of
  links $L=cN(N-1)/2$ between vertices.
  We were interested only in the connected configurations,  presented schematically in Fig. \ref{network}.
Each link in such graphs was treated as a Gaussian polymer chain with a number of monomers  $l$,
and each vertex with degree $k>1$ as a junction point, so that the resulting graph contained $M=N+L\times l$ vertices.

  We  evaluated  the size and shape properties of the set of resulting Gaussian polymer structures both
  numerically, applying Wei's
  method  and analytically, within the frames of the continuous chain model. 
  Wei's method is applicable for analysis of connected network of any (complex) topology, if  the
  Kirchhoff matrix and its eigenvalues are defined.
  The continuous chain model allows to obtain exact results of Gaussian polymer structures, but it encounters technical
  limitations
 with  increasing the complexity of polymer network architecture. Thus, the latter method have been applied in the paper only
 to some simple structures depicted in Fig. \ref{diagramy6}.
 An advantage of application of both approaches allows us to confirm the accuracy of results obtained by  Wei's method for the cases, where exact results are still unknown. 
  Our results for asphericity $\langle A_3 \rangle$ and size ratio $g$
  of polymer structures with different connectedness  $c$ are presented in  Tables \ref{table1}, \ref{table2}, \ref{table3}.
  We note, that the averaged asphericity  increases with decreasing the connectedness $c $ of corresponding graphs.
  The size ratio increases with decreasing parameter $c$: at large $c$ values, the polymer structures are more compact, whereas at smaller $c$ they become more elongated.

In forthcoming studies, we plan to use the suggested model to analyze polymer networks with higher numbers of branching points.

\section{Appendix}

\begin{figure}[t!]
	\begin{center}
		\includegraphics[width=60mm]{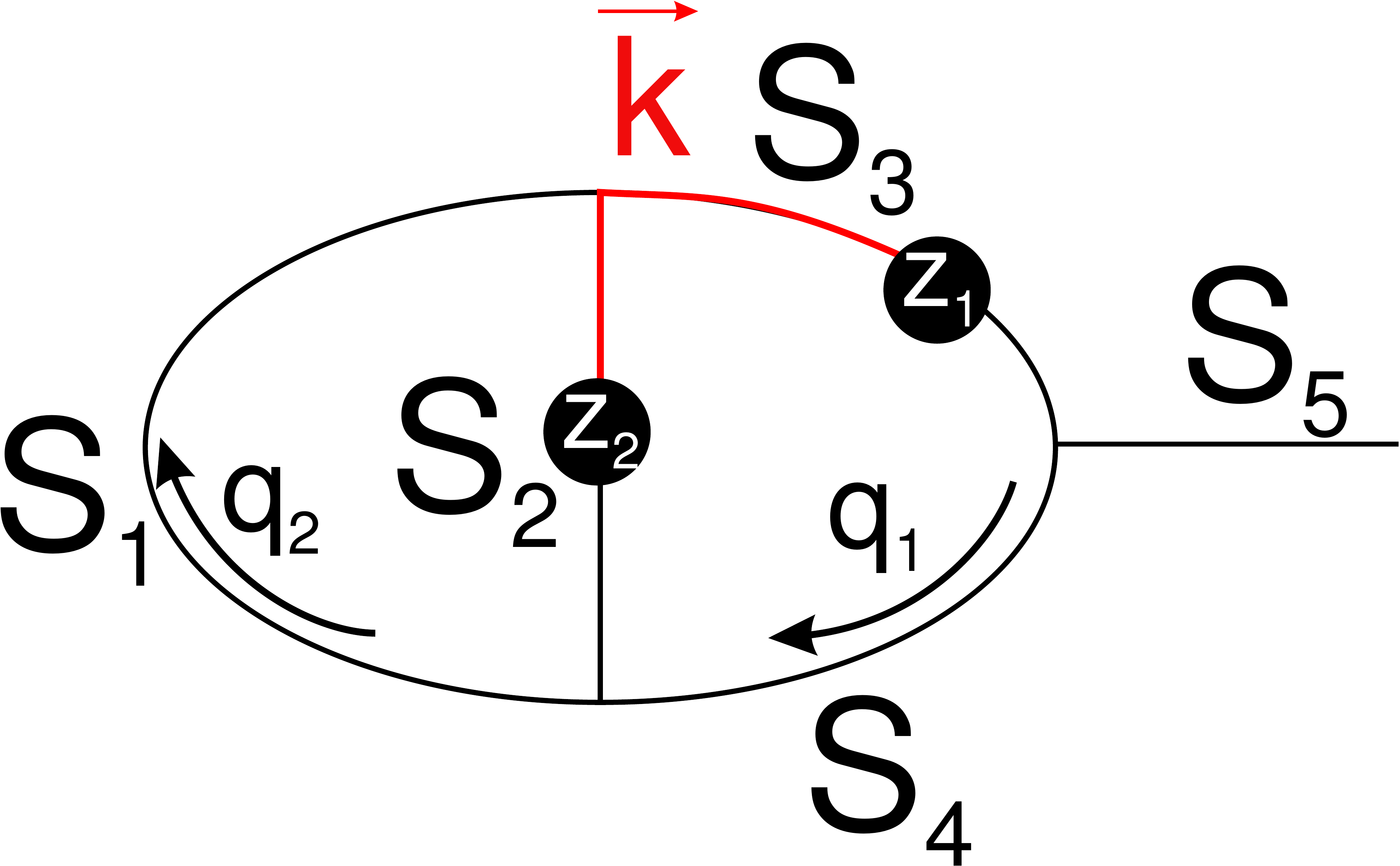}
		\caption{ \label{ED} Example of diagrammatic contribution into the gyration radius of polymer structure
			(3) at $c=0.6$, see Fig. \ref{diagramy6}.}
	\end{center}
\end{figure}

Here, we give details of calculation of the expressions, that contribute to  $\langle \xi(\vec{k}) \rangle$  at  $c=0.6$. We consider in more details diagram (d) as given in Fig. \ref{rgdiagram}. The diagram is presented in Fig. \ref{ED}. Here, the bullets  represent positions of restriction points $z_1$, $z_2$ and the red line shows the area which bears the wave  vector $\vec{k}$.  According to the rules of diagrammatic technique \cite{desCloiseaux}, this area contributes a factor $\frac{\vec{q_x}^2}{2}(x_1-x_2)$ where $\vec{q_x}$ is a sum of all vectors that pass between points $x_1$ and $x_2$ on a polymer line. Wave vectors $\vec{q}_1$, $\vec{q}_2$ appear due to the Fourier transform of $\delta$-functions, corresponding to two closed loop structures.
The analytic expression corresponding to the diagram thus reads:
\begin{eqnarray}
&&D(z_1,z_2,\vec{k})=\frac{1}{Z_6}\int \, \vec{q_1}\int \, \vec{q_2} \exp\left(-\frac{\vec{q}_1^2}{2}z_1-\frac{\vec{q}_2^2}{2}(S_4)-\right.\nonumber\\
&&\left.+\frac{(\vec{k}+\vec{q}_1)^2}{2}(S_2-z_1)-\right)\nonumber\\
&&\left(\frac{(\vec{k}+\vec{q}_1+\vec{q}_2)^2}{2}(z_2)-\frac{(\vec{q}_1+\vec{q}_2)^2}{2}(S_3-z_2)\right).
\end{eqnarray}

Integration over the wave vectors $\vec{q}_1$, $\vec{q}_2$  leads to:
\begin{eqnarray}
&&D(z_1,z_2,\vec{k})=\exp\left(-\frac{\vec{k}^2}{2}\left(S_2-z_1+z_2-\frac{(S_2-z_1+z_2)^2}{S_2+S_3}-\right.\right.\nonumber\\
&&-\left.\left.\frac{(z_2S_2+S_3z_1-S_3S_2)^2}{(S_2S_3+S_2S_4+S_3S_4)(S_2+S_3)}\right)\right)
\end{eqnarray}

Taking derivative over $|k|^2$ according to (\ref{def}) gives us
\begin{eqnarray}
&&D(z_1,z_2)= S_2-z_1+z_2-\frac{(S_2-z_1+z_2)^2}{S_2+S_3}-\nonumber\\
&&-\frac{(z_2S_2+S_3z_1-S_3S_2)^2}{(S_2S_3+S_2S_4+S_3S_4)(S_2+S_3)}.
\end{eqnarray}
The resulting contribution of the present digram into the expression for the gyration radius is obtained by integration over positions of the restriction points $z_1$, $z_2$ according to (\ref{rg}). The resulting expression reads:
\begin{eqnarray}
D=\frac{1}{6}\frac{(S_2^2S_3+S_2^2S_4+S_2S_3^2+3S_2S_3S_4+S_3^2S_4)S_3S_2}{S_2S_3+S_2S_4+S_3S_4}.
\end{eqnarray}

\section*{Acknowledgments}
We acknowledge useful discussions with Christian von Ferber and Petro Sarkanych.


\section*{References}
\begin{thebibliography}{50}
	
	
	
	\bibitem{Plaxco}
	 Plaxco KW,  Simons KT, and  Baker D (1998) J. Mol. Biol. {\bf 277} 985
	
	\bibitem{Quyang}
	Quyang Z  and  Liang J	(2008) Protein Sci. {\bf 17}  1256
	
	
	
	\bibitem{Torre01}
	de la Torre G ,   Llorca O,   Carrascosa J L,  and  Valpuesta J M (2001) Eur. Biophys. J. {\bf 30}  457
	
	
	\bibitem{Kuhn34}
	Kuhn W (1934) Kolloid-Z. {\bf 68}  2
	
	\bibitem{theory}
	 Herzog R O,   Illig R and  Kudar H (1934) Z.  Physik. Chem. Abt. A  {\bf 167}  329;
	 Perrin F (1936) J. Physiq. Radium  {\bf 7}  1
	
	\bibitem{Aronovitz86}
	 Aronovitz J A and Nelson  D R (1986) J. Physique {\bf 47}  1445 
	
	\bibitem{Rudnick86}
Rudnick	J  and  Gaspari G (1986) J. Phys. A {\bf 19}  L191
	
	
	\bibitem{Zimm49}
	Zimm H  and Stockmayer  W H (1949) J. Chem. Phys. {\bf 17}  1301
	
	
	
	\bibitem{Bishop88}
	 Bishop M and  Saltiel C J (1988) J. Chern. Phys. {\bf 88}   3976
	
	\bibitem{Diehl89}
	 Diehl H W and  Eisenriegler E (1989) J. Phys. A: Math. Gen. {\bf 22 } L87
		
	\bibitem{Gaspari87}
	 Gasparit G,   Rudnick J,  and  Beldjenna A (1987) J. Phys. A: Math. Gen. {\bf 20}  3393
	
	
	\bibitem{Honeycutt88}
	 Honeycutt J D and  Thirumalai D (1988)  J. Chem. Phys. {\bf 90}  4542
	
	\bibitem{Cannon91}
	 Cannon J W,   Aronovitz J A,  and  Goldbart P (1991) J. Phys. I (France) {\bf 1}  629
	
	
	\bibitem{Sciutto94}
	Sciutto S J  (1994) J. Phys. A Math. Gen. {\bf 27} 7015
	
	\bibitem{Haber00}
 Haber	C,   Ruiz S A, and Wirtz D (2000) PNAS   {\bf 97} 10792
	
	\bibitem{Jagodzinski92}
	 Jagodzinski O,   Eisenriegler E,  and  Kremer K (1992) J. Phys. I (France) {\bf 2}  2243
	
	\bibitem{Bishop93}
 Bishop	M,   Clarke J H R,  and  Freire J J(1993)  J. Chem. Phys. {\bf 98}  3452
	
	\bibitem{Wei97}
	Wei G (1997) Macromolecules  {\bf 30} 2125
	
	
	\bibitem{Casassa}
	Casassa F  and  Berry G C (1966) J. Polym. Sci.  Part A-2 {\bf 4}  881
	
	\bibitem{Ferber13}
	 von Ferber C,   Bishop M,   Forzaglia T,  and  Reid C (2013) Macromolecules {\bf 46}
	2468
	
	\bibitem{Ferber15}
	 von Ferber C,  Bishop M,  Forzaglia T,   Reid C,  and  Zajac G (2015) J. Chem. Phys. {\bf 142}  024901
	
	
	
	\bibitem{Duplantier89}
 Duplantier 	B (1989) J.  Stat. Phys.  {\bf 54}  581
	
	\bibitem{Schafer92}
Sch\"afer L, von Berber C, Lehr U, and Duplantier B (1992) Nucl. Phys. B {\bf 374} 473
	
	\bibitem{Ferber97}
	von Ferber C and Holovath Yu (1997) Phys. Rev. E {\bf 56} 6370
	
	\bibitem{Blavatska11}
	Blavatska V, von Ferber C, and Holovatch Yu (2011) Phys. Rev. E {\bf 83} 011803
	
	\bibitem{Gao04}
	Gao C and   Yan D (2004) Prog. Polym. Sci. {\bf 29}  183
	
	\bibitem{Jeon18}
	 Jeon I-Y,   Noh H-J,  and  Baek J-B (2018) Molecules  {\bf 23}  657
	
	\bibitem{Gu19}
	 Gu Y,    Zhao J,  and Johnson J (2019) Angew. Chem. Int. Ed.  10.1002/anie.201902900
	
	
	
	
	\bibitem{Li16}
	 Li J and  Mooney D J (2016) Nat. Rev. Mater. {\bf 1}  16071
	
	\bibitem{Lee01}
	 Lee K Y and   Mooney D J (2001) Chem. Rev.   {\bf 101}  1869.
	
	
	\bibitem{Zhou10}
	 Zhou Y,   Huang W,   Liu J,   Zhu X,   and  Yan D (2010) Adv. Mater.  {\bf 22}  4567
	
	\bibitem{McKeown06}
	 McKeown N B and  Budd P M(2006) Chem. Soc. Rev.   {\bf 35}  675
	
	
	
	\bibitem{Dusek69}
	 Du\v{s}ek K and  Prins W (1969) Adv. Polym. Sci   {\bf 6}  102.
	
	\bibitem{Clark87}
	 Clark A H and  Ross-Murphy S B (1987)  Adv. Polym. Sci.   {\bf 83}  57.
	
	\bibitem{Burchard90}
	Burchard W and  Ross-Murphy S B (eds.) 1990 {\it Physical Networks  Polymers and Gels} ( Elsevier:  Barking)
	
	
	\bibitem{Winnik97}
	Winnik M A and  Yekta A (1997)  Curr. Opin. Colloid Interface Sci. {\bf 2} 424
	
	\bibitem{Chassenieux11}
	Chassenieux C,  Nicolai T, and  Benyahia L (2011)  Curr. Opin. Colloid Interface Sci.
	{\bf 16} 18
	
	
	
	\bibitem{Dusek12}
	  Du\v{s}ek  K and  Du\v{s}kova-Smr\v{c}kov\`a M (2012) Macromol. React. Eng.  {\bf  6}  426
	
	\bibitem{Dolgushev17}
	 Galiceanu M,   de Carvalho L T,   M\"{u}lken O,  and  Dolgushev  M (2017) Polymers  {\bf 11}  577
	
	\bibitem{Dolgushev18}
	 Jurjiu A,   Maia J\"nior D G, and Galiceanu M (2018) Sci. Rep.   {\bf 8}  3731
	
	\bibitem{Renyi}
	Erd\"os P and  R\'enyi A  (1960) Publications of the Mathematical Institute of the Hungarian Academy of Sciences     5  17
	
	\bibitem{Wei}
	 Wei G  (1995) Physica A {\bf 222}  152; (1995) Physica A {\bf 222}  155
	
	
	\bibitem{Edwards}
	 Edwards S F (1965) Proc. Phys. Soc. Lond. {\bf 85}  613; (1965) Proc. Phys. Soc. Lond. {\bf 88}  265
	
	\bibitem{Solc71}
	 Solc K and  Stockmayer W H (1971) J. Chem. Phys. {\bf 54}  2756;
	 Solc K (1971) J. Chem. Phys. {\bf 55}  335
	
	
	\bibitem{Blavatska15}
	 Blavatska V and  Metzler R (2015) J. Phys. A: Math. Theor. {\bf 48}  135001
	
	\bibitem{Gaspari}
	 Gaspari G,  Rudnick J,  and  Beldjenna A (1987) J. Phys. A {\bf 20}  3393
	
	\bibitem{deGennes}
	 de Gennes P G 1979 {\it Scaling Concepts in Polymer Physics} (Ithaca, NY: Cornell University Press)
	
	\bibitem{desCloiseaux}  des Cloizeaux J and Jannink G 1990 {\it Polymers in Solutions: Their Modelling and Structure
} (Oxford: Clarendon Press)

\bibitem{Duplantier94}  Duplantier B (1994) Nucl. Phys. B {\bf 430}  489 
	
	
	
\end{thebibliography}
\end{document}